\documentclass{article}

\usepackage{arxiv}

\usepackage[utf8]{inputenc} 
\usepackage[T1]{fontenc}    
\usepackage{hyperref}       
\usepackage{url}            
\usepackage{booktabs}       
\usepackage{amsfonts}       
\usepackage{nicefrac}       
\usepackage{microtype}      
\usepackage{lipsum}		
\usepackage{graphicx}
\usepackage[square, sort, numbers]{natbib}
\usepackage{doi}
\usepackage{color}
\usepackage[dvipsnames]{xcolor}
\usepackage{setspace}
\usepackage{authblk}

\newcommand{\added}[1]{\textcolor{black}{#1}}

\usepackage{xcolor}
\hypersetup{
    colorlinks,
    linkcolor={blue}, 
    citecolor={blue}, 
    urlcolor={blue} 
}

\title{\bf Incongruent Melting and Phase Diagram of SiC \\from Machine Learning Molecular Dynamics}

\author[$\dagger$,1]{Yu Xie\thanks{Equal contribution. Yu Xie conducted this work at Harvard before joining Microsoft.}\protect\phantom{\footnotesize 1}\textsuperscript{,}}
\author[*,1]{Menghang Wang}
\author[2,3]{Senja Ramakers}
\author[1]{Frans Spaepen}
\author[1,4]{Boris Kozinsky\thanks{Corresponding to: yuxie1@microsoft.com, bkoz@seas.harvard.edu}\protect\phantom{\footnotesize 1}\textsuperscript{,}}

\affil[1]{Harvard John A. Paulson School of Engineering and Applied Sciences}
\affil[2]{Ruhr-Universit\"at Bochum}
\affil[3]{Robert Bosch GmbH}
\affil[4]{Robert Bosch LLC}

\date{}

\begin{document}

\large

\maketitle

\doublespacing
 
\begin{abstract}
Silicon carbide (SiC) is an important technological material, but its high-temperature phase diagram has remained unclear due to conflicting experimental results about congruent versus incongruent melting. Here, we employ large-scale machine learning molecular dynamics (MLMD) simulations to gain insights into SiC decomposition and phase transitions. Our approach relies on a Bayesian active learning workflow to efficiently train an accurate machine learning force field on density functional theory data. Our large-scale simulations provide direct indication that melting of SiC proceeds incongruently via decomposition into silicon-rich and carbon phases at high temperature and pressure. During cooling at high pressures, carbon nanoclusters nucleate and grow within the homogeneous molten liquid. During heating, the decomposed mixture reversibly transitions back into a homogeneous SiC liquid. The full pressure-temperature phase diagram of SiC is systematically constructed using MLMD simulations, providing new understanding of the nature of phases, resolving long-standing inconsistencies from previous experiments and yielding technologically relevant implications for processing and deposition of this material.
\end{abstract}


\section{Introduction}

Silicon carbide (SiC) is an important technological material valued for its high hardness, mechanical strength, high thermal conductivity, and wide band gap \cite{madar2004silicon}. 
The high-temperature and high-pressure behavior of SiC is of great scientific interest for understanding planetary interiors and stellar processes, due to its identification from adsorption spectroscopy of carbon-rich extrasolar planets \cite{madhusudhan2012possible}.
Understanding the properties of SiC under extreme conditions is also crucial for various applications, including nuclear reactors \cite{katoh2019silicon}, epitaxial deposition growth of SiC \cite{matsunami1997step,yi2007electron}, and graphene synthesis \cite{zhang2020atomistic} for electronics and quantum devices.

The melting and decomposition of SiC have been investigated by experimental and computational methods for decades. 
However, various experiments have reached inconsistent conclusions and thereby raised confusion about the nature of melting of SiC.
Experiments at high temperatures and pressures are challenging because it is difficult to directly observe the kinetics of melting, and factors such as sample size and purity, and thermal gradients causing temperature inhomogeneity can further complicate interpretations. 
\added{Moreover, internal inconsistencies within individual studies and substantial disagreement between independent measurements indicate that the experimental picture remains under-established.}

In some studies, incongruent melting has been reported, with the observation that SiC decomposes into a silicon rich liquid and solid carbon upon heating \cite{ekimov2004high,togaya1998melting,daviau2017decomposition,bhaumik1996modified}.
\added{Reported incongruent melting (decomposition) onset temperatures span a broad range: multiple sets of measurements near 10 GPa clusters in the 2800--3500 K range \cite{bhaumik1996modified,bhaumik2000synthesis,dolloff1961research,togaya1998melting,ekimov2004high}, whereas a high-pressure, laser-heated diamond-anvil cell (DAC) study \cite{daviau2017decomposition} reports substantially lower decomposition temperatures ($\sim 2000$ K), creating a discontinuity with the lower-pressure results. This dispersion, compounded by conflicting reports of congruent melting \cite{hall1956high,sokolov2012melting}, highlights that the SiC decomposition phase boundary remains experimentally under-determined and motivates a qualitative and quantitative re-examination.}

Molecular dynamics (MD) simulations using density functional theory (\textit{ab initio} MD, or AIMD) and various empirical potentials were employed to investigate the melting process of SiC with computational simulations. 
\added{AIMD studies \cite{finocchi1992microscopic,saiz2020ab} do not reach extended Si-rich / C-rich phase separation, although at the highest reported temperature (11000\,K) the C–C peak in the radial distribution function (RDF) begins to overtake the Si–C peak, indicating substantial carbon clustering and reduction of silicon-carbon mixing. This captures the qualitative tendency toward decomposition, but the need for unrealistically high temperature together with limited cell size and timescale prevents quantitative determination of the true decomposition boundary or formation of fully phase-separated domains.}
On the other hand, empirical potentials, such as the Vashishta potential \cite{vashishta_interaction_2007}, the Tersoff potential \cite{tersoff1989modeling,tersoff1994chemical,yan2013melting}, and the Gao-Weber potential \cite{devanathan2007atomistic}, provide differing descriptions of uncertain accuracy. For example, simulations using the Vashishta potential show no formation of C-C bonds in either liquid or amorphous SiC \added{(see Supplementary Section 6.1)}.
Similarly, the Tersoff potentials \cite{tersoff1989modeling,tersoff1994chemical}, the Gao-Weber potential \cite{devanathan2007atomistic}, and a recent Behler-Parrinello neural network potential \cite{kubo2021machine} studied the melting of SiC at high temperature and low to moderate pressure (0--10 GPa), showing RDF characteristics where Si-C bonds dominate, with minor C-C bonds signatures.
\added{More recent SiC ML interatomic potentials target relevant phenomena, including a UF3 model for high-temperature surface sublimation at near-ambient pressures \cite{macisaac2024genetic} and a deep learning potential optimized for irradiation damage cascades \cite{liu2024deep}. But they are not designed or validated for the extended high-pressure ($>10$ GPa) solid–liquid decomposition transformations.}
Therefore, the decomposition and phase separation in the amorphous phase of SiC and its melting process remain underexplored and inconsistent across \textit{ab initio}, empirical, and machine learning atomistic simulations, as well as experimental observations.

To overcome the limitations of previous studies, we employ a machine learning force field (MLFF) trained on density functional theory (DFT) data. 
First, we collect DFT data for different SiC phases and train an MLFF to describe atomic interactions using a hierarchical Bayesian active learning workflow. This approach efficiently explores the phase space across different temperatures, pressures, and compositions.
Then, we perform large-scale molecular dynamics simulations, enabled by GPU acceleration, to reliably model the melting and decomposition of SiC at a sufficient scale. By conducting two-phase coexistence simulations, where the interface between two phases of interest is created and equilibrated, we quantitatively determine, for the first time, the temperatures and pressures of transitions between phases. 

\added{Our simulations provide strong evidence for incongruent melting at high pressure and offer insights into the mechanism of SiC decomposition into Si and C rich phases.}
The transition temperatures between the zinc-blende SiC, decomposed incongruent mixture, and the homogeneous liquid phase are identified by our simulations, at pressures ranging from 10 GPa to 120 GPa. 
We obtain the pressure-temperature phase diagram depicting the stability regions of different phases. 
Notably, our findings of the decomposition indicate that amorphous SiC can only be produced through irradiation, and not via melt-quench processes \cite{ishimaru2003electron}.
Our atomic-level results provide microscopic insights into the decomposition and melting behavior of this technologically important material, resolving discrepancies in previous experimental and computational studies.

\section{Results}
\label{sec:results}

\subsection{\added{Bayesian active learning of machine learning force field} \label{subsec:active_learning}}
\begin{figure}[htbp]
    \centering
    \includegraphics[width=\textwidth]{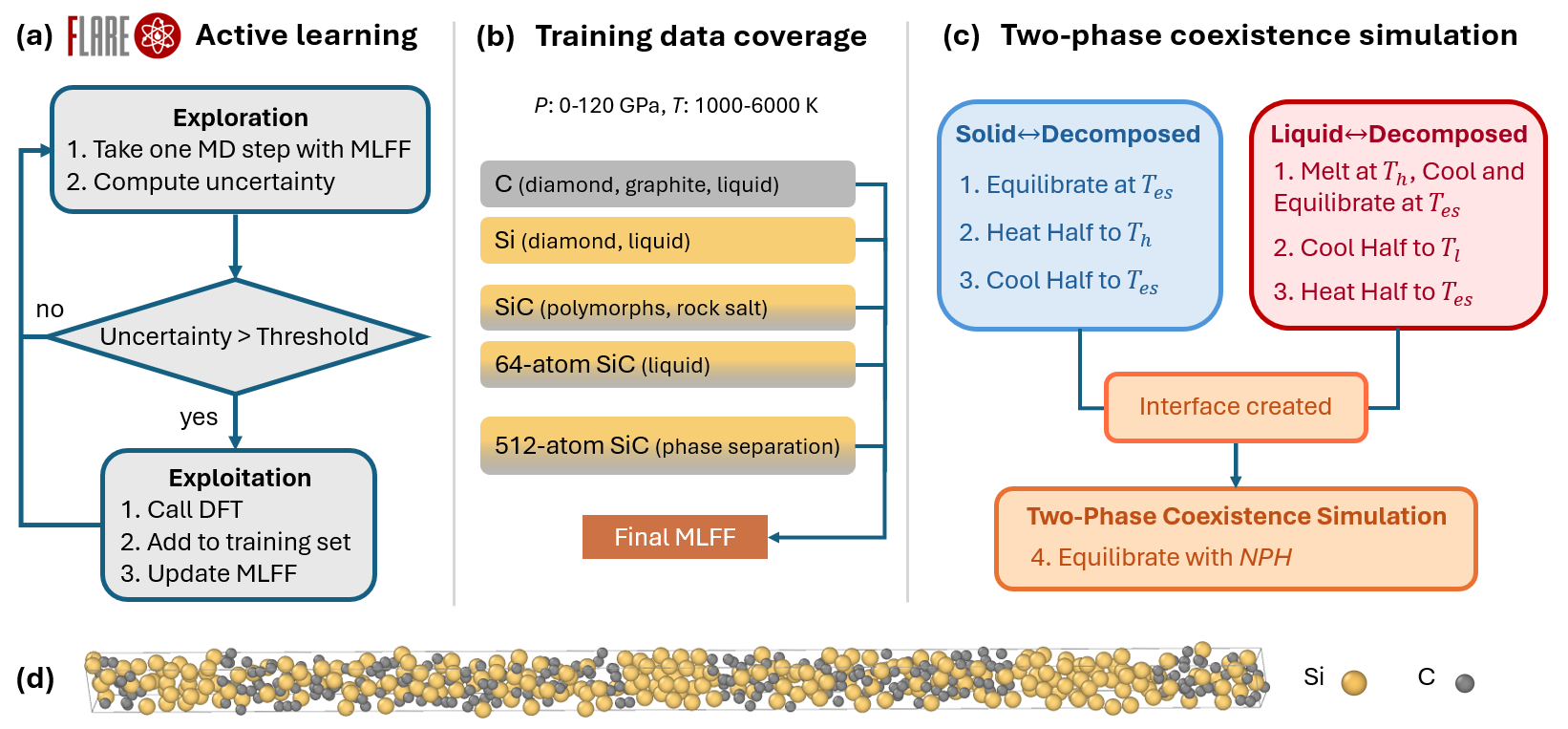}
    \caption{\doublespacing \added{a. Bayesian active learning workflow for collecting DFT training data on the fly during MD simulations. The MLFF predicts forces and uncertainties; configurations exceeding the uncertainty threshold trigger DFT labeling. 
    b. Training data composition and coverage. Multiple Bayesian active learning trajectories were performed in parallel for different compositions (pure Si, pure C, and SiC) across various temperatures and pressures. For SiC, we started with a 64-atom supercell and then extended to a 512-atom supercell where phase separation spontaneouly occurred.
    c. Two-phase simulation protocol to determine phase transition temperatures. Interfaces between the solid/liquid and decomposed phases were created by inducing decomposition, followed by a two-phase coexistence simulation in the \textit{NPH} ensemble to converge on the final transition temperature. 
    d. Spontaneous decomposition during active learning. The collected data includes states where SiC decomposition occurred spontaneously during the active learning MD simulations, prior to any decomposed structures being present in the training set.}}
    \label{fig:workflow}
\end{figure}

\added{Our MLFF is based on Gaussian process regression, which provides both force predictions and per-atom uncertainty estimates \cite{vandermause2020fly,xie2021bayesian}. This enables a Bayesian active learning workflow (Fig.~\ref{fig:workflow}a) where MD simulations are propagated with the current surrogate, and configurations whose uncertainties exceed a threshold trigger single-point DFT queries for training data collection and surrogate update.}
As more details described in the \textit{Methods} section, the MLFF was iteratively refined via Bayesian active learning to ensure accuracy across the phase space, while allowing us to simulate large systems efficiently.
\added{We performed multiple Bayesian active learning trajectories in parallel for different compositions (pure Si, pure C, and SiC) at various temperatures and pressures (Fig.~\ref{fig:workflow}b).}

While a 64-atom cell is insufficient to capture decomposition, the larger 512-atom supercell enables observation of the phase separation process.
\added{As shown in Fig.~\ref{fig:workflow}d, the first Si-rich and C-rich domains emerged spontaneously at 30 GPa during unbiased high-$T$ / high-$P$ active learning MD, prior to any decomposed structures being present in the training set. 
We did not seed or impose this Si and C phase separation. Instead, trajectories were advanced with the current surrogate, and single-point DFT labels were added only when Bayesian force uncertainties on naturally visited clustered configurations exceeded the acquisition threshold. This procedure preserves high-fidelity force predictions on-the-fly and expands the training set into newly accessed regions of configuration space.
Consequently, the database expanded to include Si-rich and C-rich regimes solely through the natural evolution of system's dynamics. This spontaneous appearance provides independent evidence that incongruent melting (decomposition) occurs without being assumed a priori, even in modest cell sizes. 
}

\added{Nevertheless, the phase separation observed in the 512-atom active learning simulations does not provide comprehensive information about the decomposition process. For instance, the limited cell size is insufficient to reveal whether the C and Si clusters form crystalline or liquid phases, or to determine the precise conditions under which decomposition occurs. To \textit{qualitatively} characterize the nature of these phases and their structural properties, we perform large-scale MD simulations with up to 512,000 atoms (Sec.~\ref{subsec:melting}). Furthermore, to \textit{quantitatively} determine the precise phase transition temperatures and pressures, we employ two-phase coexistence simulations (Sec.~\ref{subsec:high_T_twophase} and Sec.~\ref{subsec:low_T_twophase}).}

\subsection{\added{Incongruent melting from large-scale MD} \label{subsec:melting}}
Starting with the well-trained MLFF model, we run large-scale heating and cooling simulations at 30, 60, and 80 GPa to \added{\textit{qualitatively} characterize the nature of} the phase transformation and decomposition of SiC. 
We start with a 512,000-atom bulk crystal supercell of the cubic zinc-blende (ZB/3C/B3) phase.
At these pressures, decomposition does not occur readily during the MD simulations due to hysteresis, even when heating the cubic ZB crystal structure to temperatures as high as 4000\,K. The crystal structure is only destroyed when the temperature is increased to 5000\,K.

\begin{figure}[htbp]
    \centering
    \includegraphics[width=\textwidth]{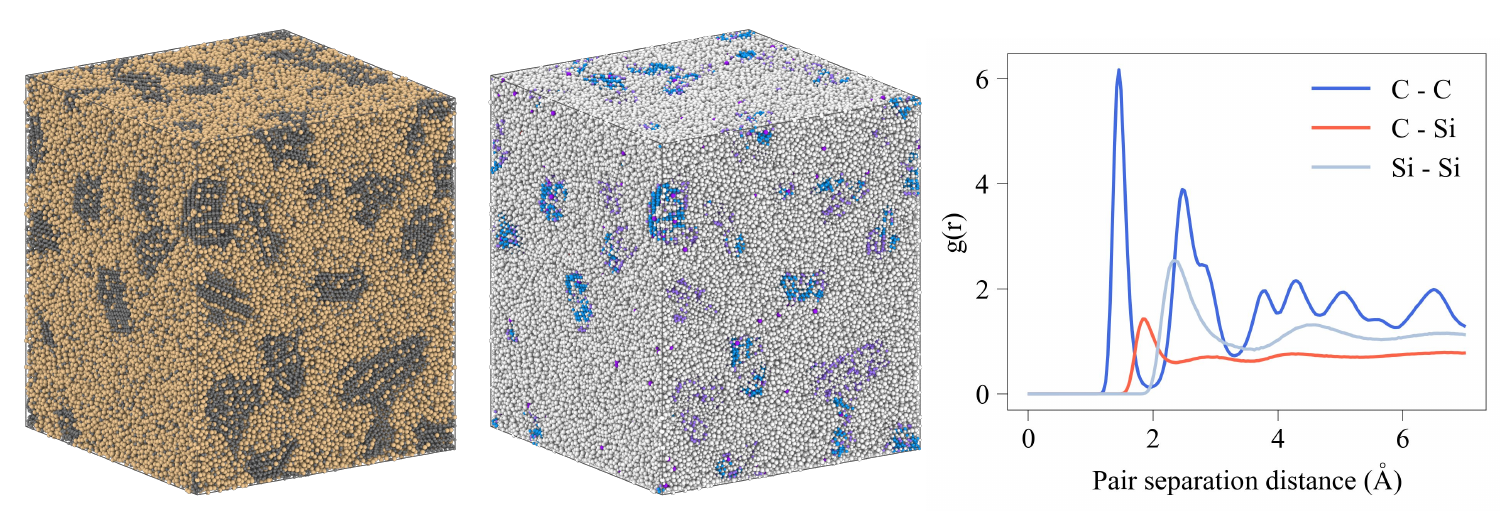}
    \caption{\doublespacing Left: decomposed configuration of 512,000 atoms at the end of the simulation at 3000\,K and 60 GPa. Yellow: Si, black: C. Middle: atoms colored with lattice type classified by polyhedral template matching. Blue: cubic diamond, orange: hexagonal diamond, purple: graphite, white: others. Right: radial distribution function of the structure \added{shows the C-C peak dominates over the Si-C peak}.}
    \label{fig:md_melt}
\end{figure}


At 5000\,K \added{across 30, 60 and 80 GPa}, small separate carbon and silicon clusters form in the amorphous configurations, which indicates a trend away from a homogeneous mixture of silicon and carbon.
Even though the temperature is significantly higher than experimentally reported incongruent or congruent melting points \cite{daviau2017decomposition,bhaumik1996modified,bhaumik2000synthesis,ekimov2004high,togaya1998melting}, the complete phase separation of C and Si does not occur.
Upon cooling the system to 3000\,K, we observe decomposition into carbon crystals and silicon liquid. 
Specifically, as the temperature drops below 4000\,K, carbon clusters start to grow in size, while the Si concentration in the liquid phase increases.
Upon further cooling to 3000\,K, larger C clusters consisting of recognizable graphite or diamond crystal structure form within the Si-rich liquid.

The formation of either graphite or diamond C phases depends on the simulation pressure.
The C clusters are classified as graphite-like or diamond-like structures by the polyhedral template matching \cite{larsen2016robust} implemented in OVITO \cite{stukowski2009visualization}. 
At 30 GPa, C clusters form graphite-like honeycomb lattice structures, while the Si-rich liquid persists throughout our simulation temperature range (3000--5000\,K), which is higher than the Si melting point (1684\,K at 0 GPa \cite{gayler1938melting}). 
At 60 GPa, the diamond and graphite-like clusters incorporate Si atoms within them, as illustrated in Fig.~\ref{fig:md_melt}. 
In addition, the C-C pair distribution function exhibits a dominant peak around 1.5 \AA, while the C-Si peak decreases.
At 80 GPa, the decomposition results in the separation of diamond C, liquid Si, and the high-pressure rock-salt (RS/B1) phase of SiC.
\added{We have included the configurations at 30 GPa and 80 GPa and more structural analysis details in Supplementary Section 6.2 and 6.3.}

Our results agree with a recent experimental observation \cite{daviau2017decomposition} that solid C and liquid Si form upon melting (i.e., incongruent melting) of SiC at high pressures.
At 120 GPa, the zinc blende to rock salt phase transition occurs, and the high pressure stabilizes the crystal structure and prevents it from amorphization, with no liquid phase appearing up to 5000\,K.

\added{In the next sections, we quantitatively determine the transition temperatures between the crystalline, decomposed, and homogeneous liquid phases using two-phase coexistence simulations, and build the P-T phase diagram of SiC.}

\subsection{\added{High-temperature phase boundary: decomposed Si+C $\leftrightarrow$ homogeneous liquid} \label{subsec:high_T_twophase}}
\begin{figure}[htbp]
    \centering
    \includegraphics[width=0.9\textwidth]{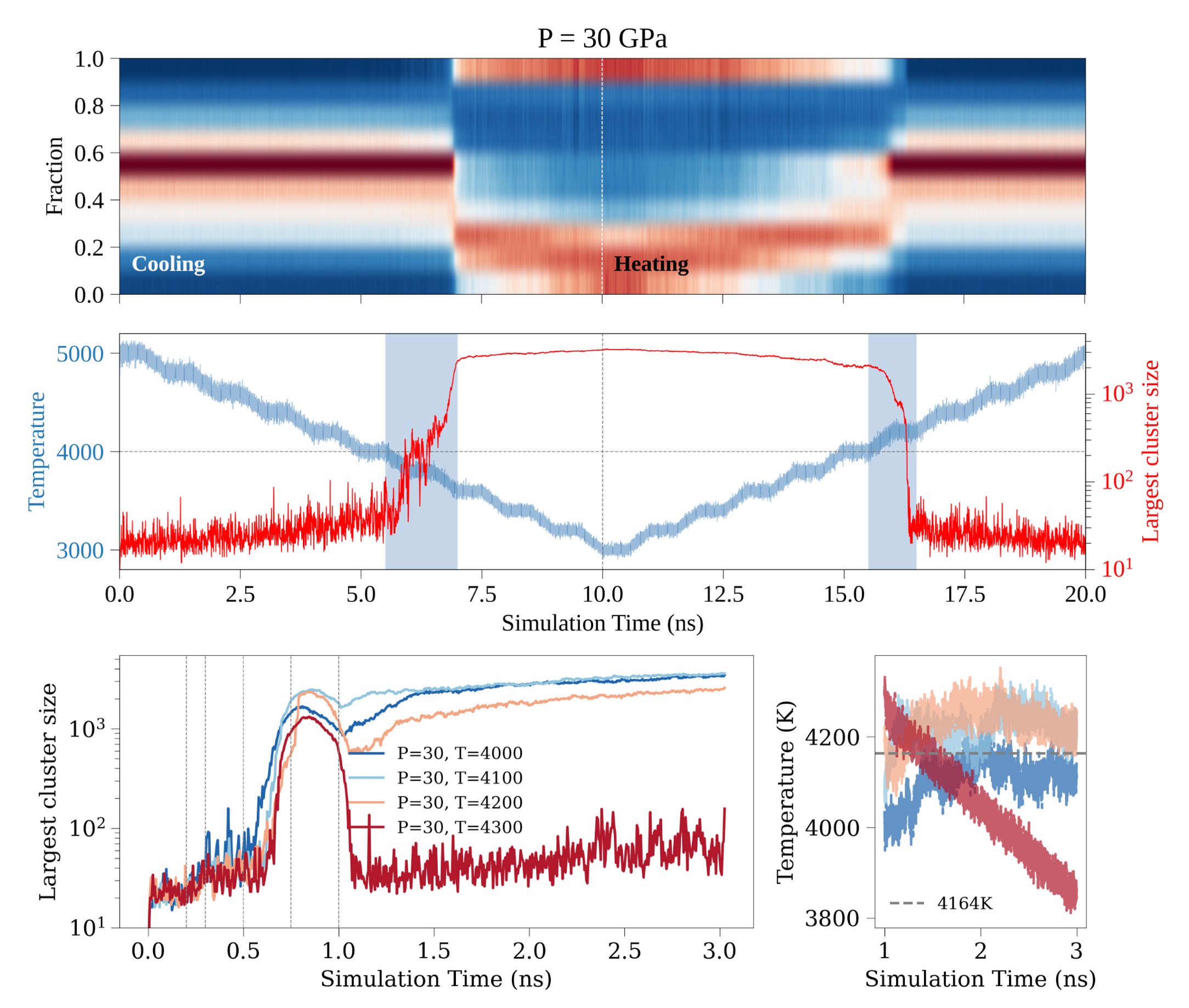}
    \caption{\doublespacing \added{SiC phase transitions and coexistence at 30 GPa from MD simulations.} Top and middle panels (from 8,000-atom cooling and heating simulations) illustrate the reversibility of phase transition. Top: population distribution of local C atom fraction over time: a single peak indicates the homogeneous liquid phase, while multiple peaks indicate the decomposed (Si+C) phase (red: high population, blue: low population). Middle: temperature (blue) and largest carbon cluster size (red, number of atoms) with the simulation time, with blue shaded regions indicating the phase transition boundaries. \added{Bottom panel (from 16,000-atom two-phase simulations) confirms the coexistence temperature by tracking the largest carbon cluster size (left) and temperature convergence at the $NPH$ stage (right).}}
    \label{fig:liquid_decompose}
\end{figure}

Having established that decomposition occurs, we next explore the phase transition between the decomposed and homogeneous liquid states.
To verify the reversibility of the incongruent high-pressure melting and to quantitatively identify the transition temperature and pressure of the spontaneous decomposition, we set up a smaller MD simulation with slower cooling and heating rates.
Specifically, we start with 8,000 atoms in the zinc-blende (B3) phase of SiC at 5000\,K, so that the crystal becomes a homogeneous liquid.
\added{A conservative simulation time step of 0.5 fs was selected due to the elevated temperatures and pressures, the short vibrational periods of light Si and C atoms, and the need to accurately capture rapid bond rearrangements during decomposition while minimizing integration error.}
We perform an equilibration for 0.5 ns at each temperature, followed by a 0.5 ns cooling process at a rate of 400 K/ns. 
We alternate between equilibration and cooling steps until reaching 3000\,K, then we ramp the temperature back up to 5000\,K following the same scheme.
The temperature profile over simulation time is shown as the blue curve in the middle panel of Fig.~\ref{fig:liquid_decompose}.

To identify spontaneous phase separation, we use the local C fraction (concentration) as the order parameter and track the size of the largest C cluster. 
At each time step, we divide the simulation cell into $10\times 10\times 10$ voxels and calculate the C atom fraction in each voxel. The population of voxels with different local fraction levels over time is represented by the color scale in the top panel of Fig.~\ref{fig:liquid_decompose}, with red indicating higher populations.

At the beginning of the simulation, at a pressure under 30 GPa and a high temperature of 5000\,K, we observe a high population of 0.5 local C fraction values, indicating a homogeneous liquid with equal mixture of Si and C atoms.
The middle panel of Fig.~\ref{fig:liquid_decompose} shows the system temperature and the largest C cluster size. As the system cools, the largest C cluster size increases, notably around 4000\,K. 
At 7 ns, when the system cools down to 3600 K, spontaneous decomposition is finished, with C atoms mostly depleted from the Si liquid. The local C-fractions are either near 0 (Si rich) or near 1 (C rich), as indicated in the upper panel.

During heating, the reverse process initiates at the same transition temperature of approximately 4000\,K. The C cluster dissolves into the Si liquid, and the local C fraction returns to around 0.5, analogous to the initial state of the simulation.
This indicates that the system has returned to the homogeneous liquid upon heating. 
Similar results are obtained for 60 GPa and 90 GPa, which are shown in the Supplementary \added{Section 7.1}.

\added{Having confirmed the reversibility, we proceed with a quantitative estimation of the transition temperature between the homogeneous liquid and decomposed Si + C phase, with a four-stage simulation protocols shown in Fig.~\ref{fig:workflow}c:
\begin{enumerate}
    \item \textbf{Initial Equilibration:} The crystalline supercell is first melted and then cooled to the estimated transition temperature $T_{es}$. The entire cell is then equilibrated at this temperature to prepare the initial liquid phase.
    \item \textbf{Interface Creation:} To create an interface, the positions and velocities of half the atoms in a slab along $z$-axis are fixed, while the other half is cooled further to a lower temperature, $T_l$ (choosen here as 3500 K) to induce the formation of the decomposed phase.
    \item \textbf{Interface Refinement:} The cooled region is heated back from $T_l$ to $T_{es}$ in the $NP_{zz}T$ ensemble. This action ensures a minimal thermal gradient across the cell while creating a stable interface between the decomposed phase and the liquid phase.
    \item \textbf{$NPH$ Coexistence:} Constraints are removed, and the entire system is equilibrated in the $NPH$ ensemble. The phase boundary equilibrates, and the system temperature converges to the phase transition temperature.
\end{enumerate}
The first three stages prepare the two-phase interface near the estimated transition point. During the final $NPH$ stage, the system equilibrates. The temperature at which both phases coexist in equilibrium is then determined as the phase transition temperature.
}

\added{As shown in bottom panels of Fig.~\ref{fig:liquid_decompose}, the initial liquid phase at $T_{es}$ is prepared after 0.5 ns (0.2 ns melting, 0.1 ns cooling, and 0.2 ns equilibration). Then, the decomposed phase is clearly formed during stage 2 and 3 (achieved by 0.25 ns of cooling followed by 0.25 ns of heating), indicated by the sharp increase in the size of the largest carbon cluster. Starting at the 1 ns mark, the system enters the final $NPH$ coexistence stage and subsequently converges to a final transition temperature of 4164\,K in the coexistence simulation. We note that for runs initialized with an estimated temperature of $T_{es}=4300K$, the temperature was too high to stablize the decomposed phase, and temperature convergence was therefore not achieved. Thus, we computed the final average transition temperature from the remaining three converged runs. Additional two-phase simulation results for pressures at 45, 60, 70 GPa are provided in the Supplementary Section 7.1.}

\subsection{\added{Low-temperature phase boundary: crystal $\leftrightarrow$ decomposed Si+C} \label{subsec:low_T_twophase}}

Next, we investigate the lower-temperature transition between the crystal (B1 or B3) phases and the decomposed phase. To overcome hysteresis and high nucleation barriers that prevent decomposition in direct MD simulations of B3 SiC crystals heated to 4000\,K, we employ \added{a similar four-stage two-phase coexistence simulation approach \cite{dozhdikov2012two}, as illustrated in Fig.~\ref{fig:workflow}c. 
\begin{enumerate}
    \item \textbf{Initial Equilibration:} Equilibrate the crystalline supercell at pressure $P$ (10-90 GPa) and an initially estimated transition temperature $T_{es}$. 
    \item \textbf{Interface Creation:} Fix half of the $z$-slab, while melt the other half by heating it to high temperature $T_h \gg T_{es}$ in the $NP_{zz}T$ ensemble to induce the liquid phase.
    \item \textbf{Interface Refinement:} The melted half is then cooled from $T_h$ to $T_{es}$ in the $NP_{zz}T$ ensemble. This process facilitates the formation of the decomposed phase while maintaining the unperturbed crystalline half, thus creating a stable crystal-decomposed interface.
    \item \textbf{$NPH$ Coexistence:} Constraints are removed, and the entire system is equilibrated in the $NPH$ ensemble. The phase boundary equilibrates, and the system temperature converges to the phase transition temperature.
\end{enumerate}
}

\begin{figure}[htbp]
    \centering
    \includegraphics[width=0.8\textwidth]{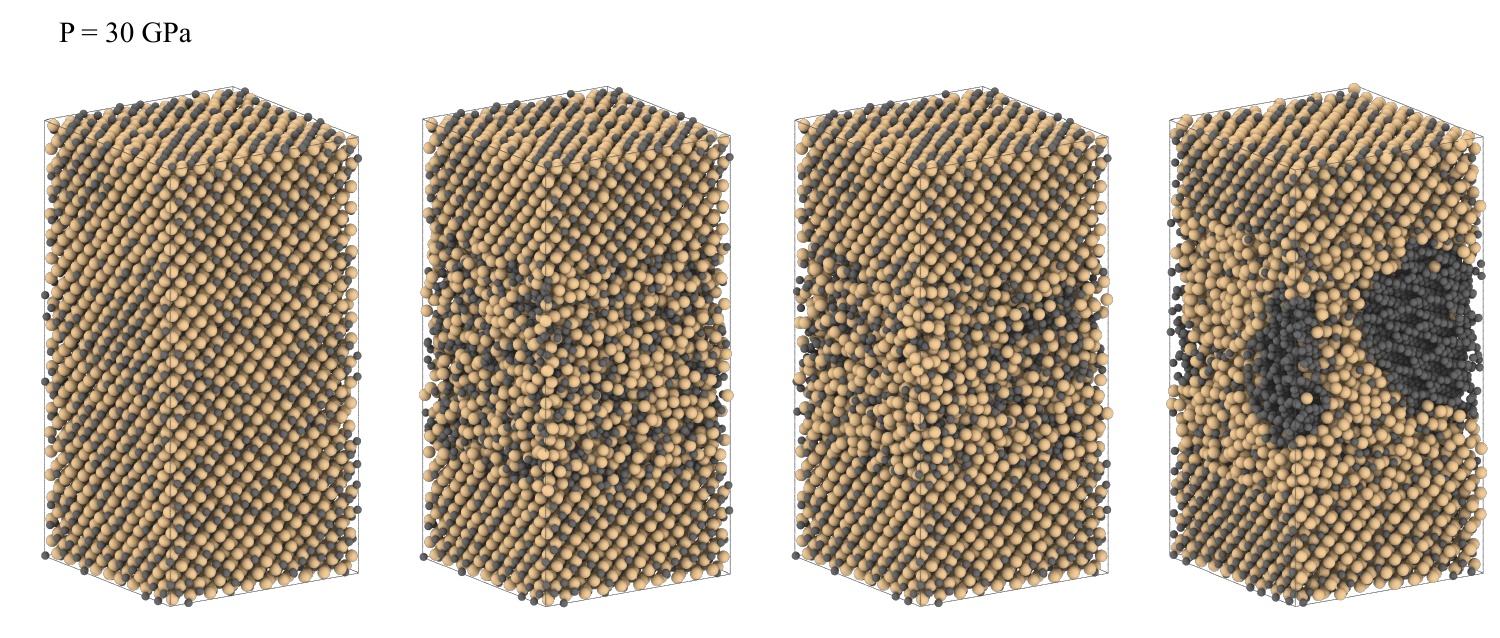}
    \includegraphics[width=0.8\textwidth]{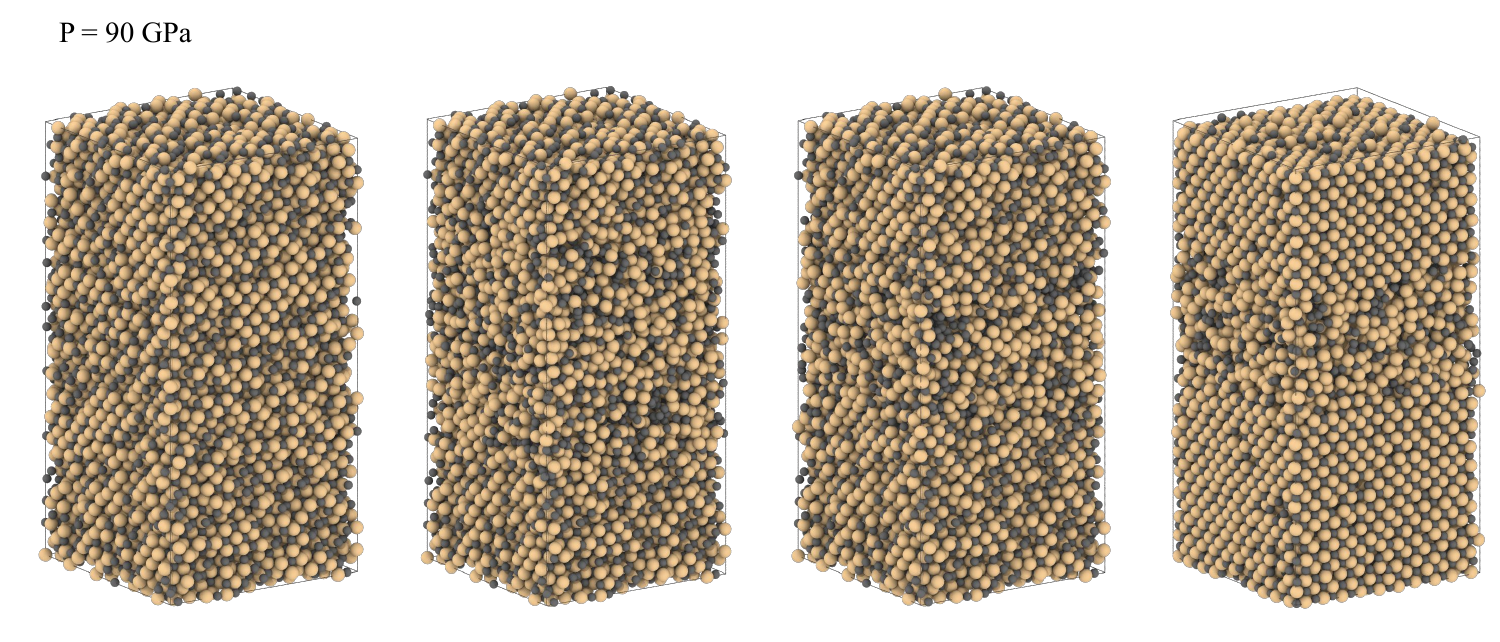}
    \caption{\doublespacing Snapshots of the four stages of the two-phase MD at 30 GPa (top) and 90 GPa (bottom) pressures. Each stage is simulated for 0.5 ns.}
    \label{fig:two-phase-snapshots}
\end{figure}

Each of the four stages is simulated for 0.5 ns, and the snapshots at the end of each stage are shown in Fig. \ref{fig:two-phase-snapshots} for the pressure values of 30 GPa (starting with B3 zinc blende) and 90 GPa (starting with B1 rock salt).
In the last stage of the two-phase simulation where the interface is allowed to move, the temperature of the entire system fluctuates and converges to the actual transition temperature at which the decomposed and crystalline phases coexist in equilibrium, and the interface stops moving.  

\begin{figure}[htbp]
    \includegraphics[width=0.9\textwidth]{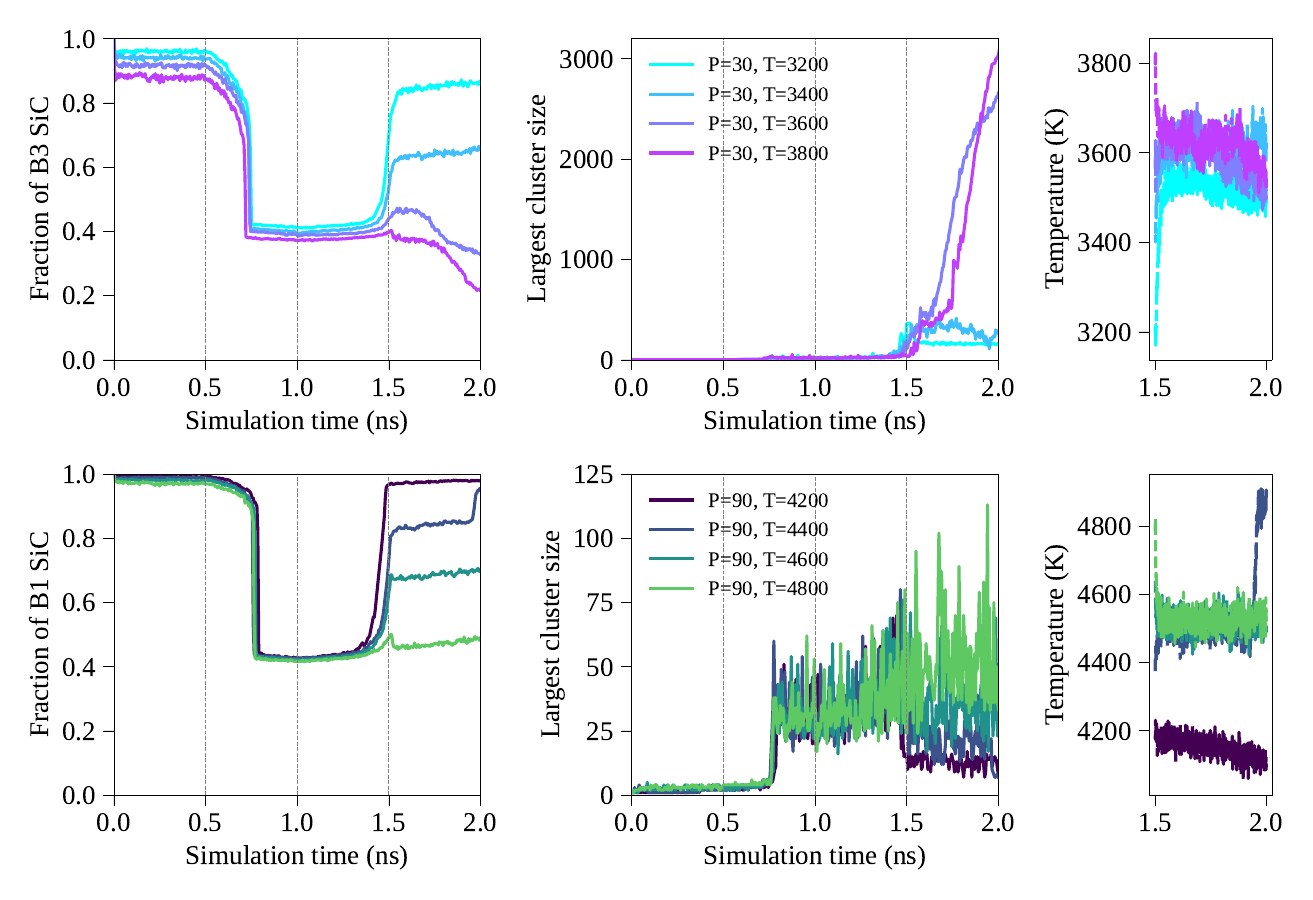}
    \caption{\doublespacing Analysis of two-phase coexistence simulations. Panels (left to right) show: fraction of zinc-blende (rock-salt) structures, largest C cluster size (number of atoms), temperature convergence at the $NPH$ stage, over simulation time. Top: 30 GPa. Bottom: 90 GPa.}
    \label{fig:two-phase-analysis}
\end{figure}
Fig.~\ref{fig:two-phase-analysis} shows the results of the heating and cooling process in the two-phase coexistence simulations for different initial temperatures near the expected transition temperature. 

At the fixed pressure of 30 GPa, during the heating stage (0.5--1 ns), the heated half of the supercell becomes amorphous, and the overall fraction of zinc-blende local structures in the supercell decreases to 0.5.
The C clusters begin to form during the cooling stage (1--1.5 ns) and continue growing in size at 1.5--2 ns as demonstrated in the second column of Fig.~\ref{fig:two-phase-analysis}. 
In the NPH simulation stage (1.5--2 ns), both B3 crystal and decomposed Si + C phases coexist, with relatively different equilibrated fractions dependent on the initial temperature. 
Meanwhile, multiple simulations with different initial temperatures converge to approximately the same temperature during the NPH stage, which is identified as the transition temperature (shown in the third column of Fig.~\ref{fig:two-phase-analysis}).

At 90 GPa, where the stable phase at low temperature is rock salt (B1), a similar procedure starting from B1 phase yields a transition temperature around 4500\,K.
Notably, the carbon cluster sizes formed at 90 GPa are much smaller than those at 30 GPa, leading us to conjecture that at sufficiently high pressures, the decomposed Si+C phase between the crystalline and liquid phases may not be present. 
\added{Additional coexistence simulations were performed for the B3 phase (at 10, 20, 45, 60, 70, and 80 GPa) and the B1 phase (at 85 and 90 GPa). In addition, we investigated finite-size effects by performing significantly larger simulations (4x and 8x the original size). These results and the details for determining the final phase transition temperatures are presented in the Supplementary Section 7.3, 7.4, and 7.6.}

\subsection{Phase diagram}

\begin{figure}[htbp]
    \centering
    \includegraphics[width=\textwidth]{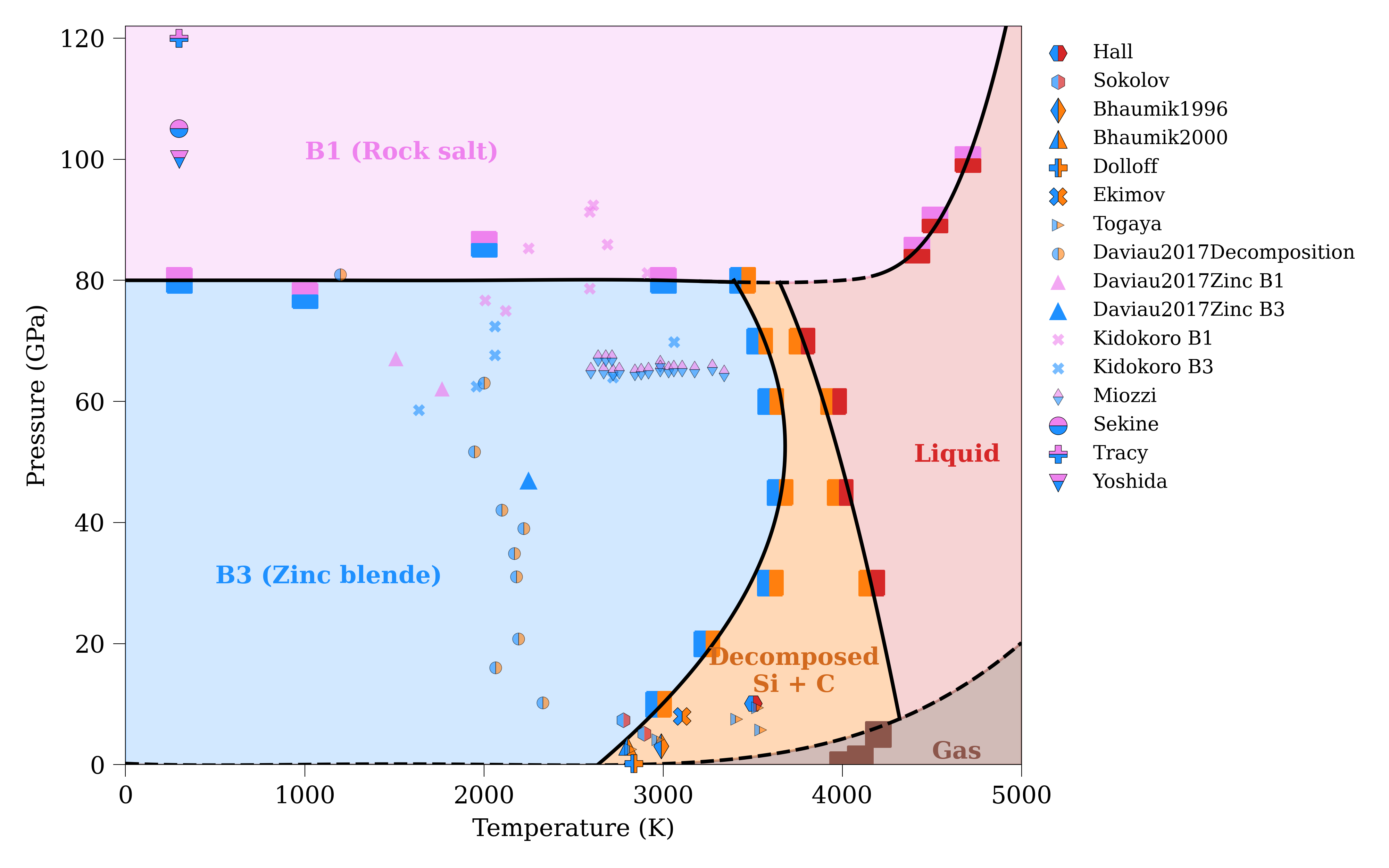}
    \caption{\doublespacing Phase diagram of SiC. Results from our MD simulations are denoted in big squares, and the inferred phase boundaries are drawn with black solid lines (interpolated) and dash lines (extrapolated). Blue: B3 (ZB), pink: B1 (RS), orange: decomposed Si + C, red: homogeneous liquid, brown: gas. 
    The background colors indicate the stability regions of different phases inferred by our simulations. 
    Experimental reports of congruent melting (blue-red symbols): hexagons \cite{hall1956high}, rotated hexagons \cite{sokolov2012melting}. 
    Incongruent melting (blue-orange symbols): diamonds \cite{bhaumik1996modified}, triangles \cite{bhaumik2000synthesis}, pluses \cite{dolloff1961research}, crosses \cite{ekimov2004high}, left-pointing triangles \cite{togaya1998melting}, circles \cite{daviau2017decomposition}. 
    B3-B1 transition (blue-pink): triangles \cite{daviau2017zinc}, crosses \cite{kidokoro2017phase}, diamonds \cite{miozzi2018equation}, circles \cite{sekine1997shock}, pluses \cite{tracy2019n}, inverted triangles \cite{yoshida1993pressure}.} 
    \label{fig:diagram}
\end{figure}

Finally, we combine the results obtained from the MD simulations of the SiC phase transitions into a complete pressure-temperature phase diagram. Fig.~\ref{fig:diagram} presents the $P$-$T$ diagram, illustrating the B3 $\leftrightarrow$ B1, B3 $\leftrightarrow$ gas, B3 $\leftrightarrow$ Si + C, B1 $\leftrightarrow$ Si + C, and Si + C $\leftrightarrow$ SiC (liquid) transitions.
For completeness, we briefly describe the simulations of B3 $\leftrightarrow$ B1 transition and B3 $\leftrightarrow$ gas below, although these are not the primary focus of this work.

The B3 $\leftrightarrow$ B1 transition pressures at various temperatures are determined using the same methods as in our previous work \cite{xie2023uncertainty}. 
Starting with the B3 crystal, we increase pressure at a fixed temperature to generate a configuration containing both phases. Then, we perform constant-pressure MD simulations on these coexistence structures at different pressures to identify the transition point, characterized by the stable coexistence of two phases remains. 
The pressure scan results are shown in the Supplementary \added{Section 8.1}.
The B3 $\leftrightarrow$ B1 transition pressures identified from our simulations (along pink-blue boundary in Fig.~\ref{fig:diagram}) are close to experimental measurements with a deviation of less than 20 GPa. 

At zero or low pressure, the SiC B3 crystal sublimates into gas phase instead of melting, which only occurs under high pressure. 
\added{Our $NPT$ MD simulations at $P$ = 0--15 GPa and $T$ = 3900--4500 K reveal a qualitative pressure threshold between 5 and 10 GPa: at lower pressures (0, 1, and 5 GPa), sustained volumetric expansion above 4000--4200 K indicates progressive sublimation to a gas/vapor phase (details in Supplementary Section 8.2). While our simulations identify this sublimation threshold, most experimental literature at low pressure focuses on sublimation growth conditions rather than on establishing the equilibrium solid-gas phase boundary.
An earlier study \cite{krieger1968thermodynamics} reported a sublimation temperature-pressure relationship, but it did not distinguish between the specific solid phases of SiC.}

\added{Regarding the transition between condensed phases, there is a qualitative disagreement among experiments: Hall \cite{hall1956high} and Sokolov \cite{sokolov2012melting} report congruent melting, whereas Daviau \textit{et al.} \cite{daviau2017decomposition} and five earlier independent studies (Bhaumik \textit{et al.} \cite{bhaumik1996modified,bhaumik2000synthesis}, Dolloff \cite{dolloff1961research}, Togaya \cite{togaya1998melting}, Ekimov \textit{et al.} \cite{ekimov2004high}) observe incongruent decomposition—consistent with our simulations. Quantitatively, at 10 GPa, our predicted decomposition temperature falls within the cluster formed by those five independent measurements, supporting our placement of the boundary there. At higher pressures, however, our values are substantially above the laser-heated DAC series from Daviau \textit{et al.} \cite{daviau2017decomposition} (blue-orange circles). This laser-heated DAC dataset shows several inconsistencies. When extrapolated downward, it sits discontinuously below the lower-pressure data. It also shows internal inconsistency with a separate study by the same group \cite{daviau2017zinc} (blue triangles). Furthermore, it conflicts with other B3-B1 studies \cite{kidokoro2017phase,miozzi2018equation} (blue cross and pink-blue diamonds) that report stable B3-SiC in a region where Daviau \textit{et al.} claimed decomposition occurs. Together with the broader scatter among datasets, this indicates that the high-pressure decomposition boundary remains experimentally underconstrained. Potential sources for these discrepancies, such as temperature gradients and transients in the DAC environment, highlight the need for further high-pressure measurements to resolve these conflicts.}


The high-temperature phase boundary is close to the melting point of graphite and diamond carbon, indicating that the C clusters dissolve at around 4000\,K and SiC becomes a homogeneous liquid mixture.
As depicted in Supplementary Fig.\added{~S3}, the Si-C peak indicates that the Si and C are mixed well, and the complete phase separation does not occur, as opposed to the RDF shown in Fig. \ref{fig:md_melt}.

\section{Discussion}


\paragraph{Uncertainty aware simulations}
Our MLFF, based on Gaussian process regression \cite{vandermause2022active}, provides uncertainty estimations for its predictions. To ensure the fidelity of our large-scale MD simulations, we assess whether high-uncertainty configurations are involved.
A sufficient training dataset covering a variety of configurations results in low uncertainties during simulations, as the model is familiar with the encountered atomic structures. Conversely, high model uncertainty implies the MD has proceeded to configurations not covered by the training data, potentially reducing prediction reliability.
Therefore, we evaluate the model uncertainty on snapshots from the decomposition MD simulations. As shown in Supplementary Figure S16, the model uncertainty remains well below the acquisition criteria of our force field active learning workflow. 
This indicates that our collected training data set sufficiently covers the configuration space, ensuring high confidence level in the model predictions.


\paragraph{Heating-cooling protocol}
In our MD simulation where decomposition occurs, we cannot directly model the transition from a pure crystal B3 phase to the decomposed phase due to hysteresis and nucleation barriers. Instead, we must overheat and amorphize the crystal SiC to a homogeneous liquid before decomposition can take place during the cooling stage.
Meanwhile, experimental observations align with this approach, demonstrating that the decomposition originates from the heating and cooling processes. For example, Daviau and Lee reported no evidence of decomposition in the Raman spectra when heating to 3200\,K at 81\,GPa. However, the surrounding region, annealed to 1200\,K, showed D and G band signatures of carbide-derived carbon \cite{daviau2017decomposition}.
Similarly, Togaya and Sugiyama \cite{togaya1998melting} reported using a sample quenched from a molten state to determine the onset of fusion, further supporting the importance of cooling in the decomposition process.

\paragraph{Grain size effects on congruent melting observations}
While numerous experiments have confirmed the occurrence of incongruent melting or decomposition in SiC, some studies have reported congruent melting. Considering the slow kinetics of SiC decomposition and phase transformation, the grain size can influence the process. For instance, in experiments reporting congruent melting, large single crystalline grains (between 150 $\mu$m and 3 mm) are used, which likely hinders the transition \cite{sokolov2012melting}.

\paragraph{Limitations of simulations and experiments}
While our MD simulations confirm incongruent melting at various pressures, qualitatively consistent with Daviau \textit{et al.} \cite{daviau2017decomposition}, our obtained transition temperatures at high pressures are significantly higher than the reported measurements around 2000\,K.
\added{
One limitation of our MLFF is that it is trained primarily on high-temperature/high-pressure data, focusing on bulk solid-solid and solid-liquid transitions, but not optimized for simulating surfaces or capturing detailed effects across a wide range of complex crystalline defect ensembles.
The expressive power of MLFF and the accuracy of DFT can contribute to the discrepancy of our simulation results with experiments. 
More expressive and expensive MLFF models such as the equivariant neural works \cite{batzner20223,musaelian2023learning} can be used to potentially improve the accuracy.
}

\added{On the other hand, the discrepancy may also} be attributed to several factors in experimental setups:
(1) Since the temperature is measured via the thermal emission from the surface, the approach could underestimate the core temperature. 
(2) The heat distribution in the sample could be inhomogeneous, such that overheating in certain regions might facilitate the decomposition at lower apparent temperatures.
Although there are no other experimental reports at such high pressures, our simulation is in good agreement with the transition temperature at a lower pressure of 10 GPa found in \added{five other experiments (Bhaumik \textit{et al.} \cite{bhaumik1996modified, bhaumik2000synthesis}, Dolloff \cite{dolloff1961research}, Togaya \cite{togaya1998melting}, and Ekimov \textit{et al.} \cite{ekimov2004high}).}
We also note that in other applications, such as physical deposition vapor growth of SiC, a temperature of $\sim$ 2500\,K is used at $\sim$ 0\,GPa, demonstrating SiC's stability at temperatures higher than 2000\,K \cite{kimoto2016bulk}. This observation further supports the plausibility of our higher predicted transition temperatures at elevated pressures. 

\paragraph{Effect of point defects and surfaces}
The two-phase coexistence MD simulation creates an interface between two different phases and thereby overcomes the nucleation barrier. It is possible that the defect concentration influences the phase transition temperature. Even state-of-the-art epitaxially grown SiC wafers have high defect densities \cite{kimoto2016bulk}.
Therefore, we performed additional simulations, in which we introduce different point defect concentrations in the crystal. As shown in Supplementary Section 7.5, the concentration of defects does not have a significant influence on the transition temperature converged by the two-phase MD simulation. Macroscopic defects such as dislocations and phase boundaries could have a greater impact than point defects but are out of scope of this study. 
\added{Besides point defects, we further note that free surfaces or high surface-to-volume ratios, absent from our periodic bulk simulations, can modify local coordination, provide heterogeneous nucleation sites, and shift the decomposition/melting temperature, and thus may contribute to remaining experiment–simulation discrepancies.}

\paragraph{Amorphous SiC formation}
While some previous computational studies using empirical and neural network potentials obtain amorphous SiC through melting and quenching \cite{devanathan2007atomistic,kubo2021machine}, our simulations show that the melting and quenching process results in the decomposition of the SiC. Instead, techniques such as irradiation are required for the amorphization of SiC, which is consistent with most of the experiments \cite{snead1998amorphization}.

\begin{figure}[h]
    \centering
    \includegraphics[width=\textwidth, trim=30mm 20mm 0mm 0mm]{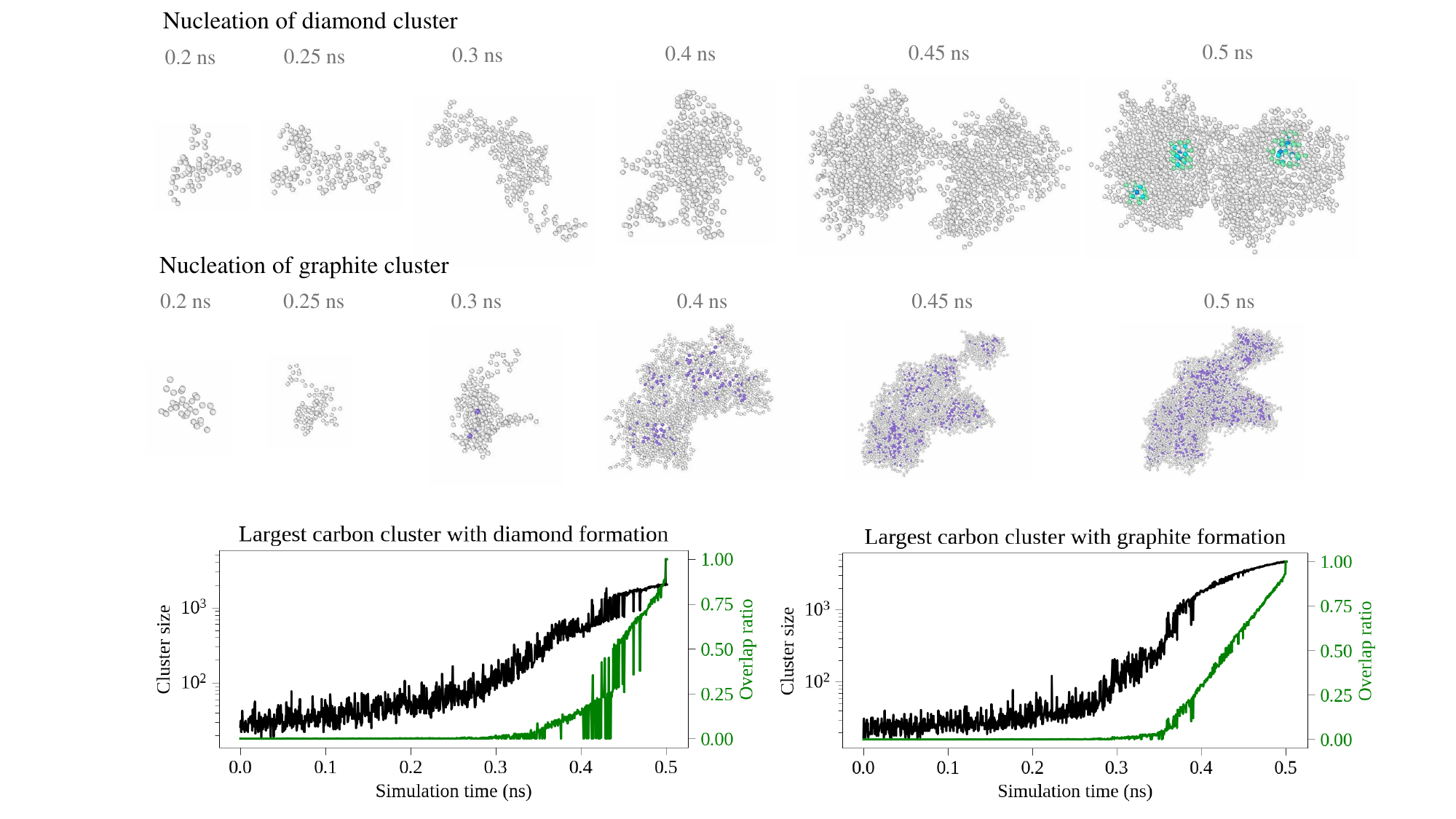}
    \vspace{5mm}
    \caption{\doublespacing Nucleation analysis of carbon clusters from simulations of 512,000 atoms. Top: Nucleation of diamond cluster at 60 GPa. Diamond core is colored in cyan. Middle: Nucleation of graphite cluster at 30 GPa. Graphite core is colored in purple. Bottom: Increase of the diamond cluster (left) and graphite cluster (right) size and its overlap with the final largest cluster, with simulation time.}
    \label{fig:carbon_nuclei}
\end{figure}

\paragraph{Decomposition mechanisms: nucleation and spinodal decomposition}
Building upon our findings on SiC decomposition, we now explore possible mechanisms underlying this process. The two primary mechanisms to consider are nucleation and spinodal decomposition. 
To investigate spinodal decomposition, we calculated the C-C spatial correlation function from local concentrations (Supplementary Section 6.5). Notably, towards the end of cooling process, particularly below 3400\,K, a peak appears at approximately 40 Å. This implies periodic fluctuations in the carbon concentration, indicative of a wavelength of the phase separation mode \cite{zhou2013quantitative,sarkar2023nucleation}. This observation aligns with the spinodal decomposition mechanism reported in alloy systems \cite{zhou2013quantitative,sarkar2023nucleation} and is consistent with a recent tight-binding MD study of SiC under tensile stresses \cite{herrero2023cubic}.
To analyze the nucleation of solid carbon clusters during decomposition, we tracked the evolution of diamond and graphite clusters.
Fig.~\ref{fig:carbon_nuclei} illustrates the various stages of carbon cluster development, highlighting their transformation into either diamond or graphite phases.
We applied cluster analysis in OVITO \cite{stukowski2009visualization} with a 1.55 Å cutoff --- approximately the bond length in graphite and diamond --- to identify clusters.
The formation of different phases were analyzed by the ``identify diamond structure'' method \cite{maras2016global} for diamond, and polyhedral template matching \cite{larsen2016robust} for graphite. 
We track the evolution of clusters by identifying those with the largest overlap with the final structures at each time step.
As shown in Fig.~\ref{fig:carbon_nuclei}, in earlier stages, small nuclei lack clear lattice symmetry or phase identity. For graphite, the hexagonal structure can be traced back to smaller clusters. However, tracking becomes challenging for clusters below 100 atoms, as dynamic rearrangements reduce the overlap ratio quickly over time for both diamond and graphite structures.

\paragraph{Conclusions}
In conclusion, our machine learning molecular dynamics study confirms the occurrence of incongruent melting in SiC, with the decomposition of SiC into Si and C being observed at a temperature range of \added{3000} -- 3600 K at high pressures. 
We also identified a transition at higher temperatures where the decomposed phase transforms into a homogeneous liquid phase. 
The highly efficient and accurate MLFF enabled previously intractable modeling and simulation of the decomposition process, overcoming limitations of both \textit{ab initio} or empirical force fields in terms of computational cost and accuracy.
Our MD simulations provide atomic-level insights, revealing nucleation mechanisms and process details inaccessible in experiments. These findings support the existence of incongruent melting at high pressure, clarifying controversial experimental observations.
In summary, our study provides a comprehensive understanding of SiC melting and decomposition behavior, along with a complete phase diagram, reconciling and extending previous experimental and theoretical work in this field.

\section{Methods}
\paragraph{Machine learning force field}
Accurately modeling the melting and decomposition of SiC at high temperature and pressure requires an interatomic potential that captures complex bonding environments, including phase separation. 
\added{Traditional empirical force fields can reproduce certain crystalline and shock-response properties. For example, Tersoff \cite{tersoff1994chemical} and Vashishta \cite{vashishta_interaction_2007} potentials have been applied in planar and impact shock studies of SiC \cite{li2018planar,branicio2018shock,li2017shock}. 
However, the concurrent bond breaking, chemical partitioning, and long-timescale phase coexistence needed to delineate the high-$P$ / high-$T$ decomposition boundary remains challenging for such potentials.}
While AIMD \cite{finocchi1992microscopic,saiz2020ab} provides accurate atomic forces, its high computational cost limits simulations to small system sizes and short timescales. To overcome these challenges, we employ a MLFF trained on DFT calculations, such that the atomic interactions can be modeled with high accuracy, while remaining cost effective for large-scale simulations.

\added{Among general MLFF frameworks \cite{Bartok2015GAP,Thompson2015SNAP,batzner20223,musaelian2023learning}, FLARE was selected because: (i) Bayesian uncertainties focus acquisition on novel crystal, liquid, and early segregation environments, enabling highly efficient collection of non-redundant data; (ii) sparse GP to polynomial mapping yields high simulation efficiency at the $10^4$--$10^6$ atom scale needed for multi-temperature/pressure coexistence runs; (iii) reuse of a previously validated SiC setup \cite{xie2023uncertainty} ensures methodological continuity. More expressive deep equivariant models could lower force errors at substantially higher cost, a trade-off we reserve for future refinements.}

Our MLFF is built using the FLARE framework \cite{vandermause2020fly,xie2023uncertainty}, which is based on a sparse Gaussian process (SGP) model \cite{vandermause2022active} with atomic cluster expansion descriptors \cite{drautz2019atomic}. This model predicts forces while providing uncertainty quantification.

\paragraph{Bayesian active learning}
To efficiently generate training data, we employ Bayesian active learning, which systematically selects new atomic configurations where the MLFF is most uncertain. The workflow follows these steps:
(1) Initial Training: The MLFF is first trained on a small set of DFT-calculated structures, including perfect and slightly perturbed crystal structures.
(2) Uncertainty-Guided MD: The MLFF runs MD simulations, predicting atomic forces while simultaneously estimating its uncertainty.
(3) Selective DFT Calculations: If the model's uncertainty exceeds a threshold, new configurations are labeled using DFT. Otherwise, continue with MD.
(4) Iterative Refinement: The MLFF is retrained with the newly acquired DFT data, progressively improving its accuracy.
This adaptive learning process ensures that the force field is only trained where necessary, avoiding redundant DFT calculations while capturing the relevant phase space.
Once the active learning is complete, we map the SGP into an equivalent but more computationally efficient polynomial model \cite{vandermause2022active,xie2023uncertainty}, which serves as the final MLFF for large-scale production simulations in LAMMPS \cite{thompson2022lammps}. 
The DFT calculations are performed using VASP \cite{kresse1993ab} with the PBE exchange correlation functional \cite{perdew1996generalized}, following the computational settings in our previous works \cite{ramakers2022effects,xie2023uncertainty}.

The initial training set contains SiC polymorphs data from our previous study \cite{xie2023uncertainty}, encompassing the hexagonal 2H, 4H, and 6H polytypes, B3 (zinc blende or 3C) and the high-pressure B1 (rock salt) phases. 
We then employ the FLARE Bayesian active learning workflow to systematically sample data that captures melting and decomposition behavior SiC across a wide pressure range.
Active learning MD simulations are performed \added{for a 64-atom SiC supercell} at 0, 30, 60, 90, and 120 GPa over a temperature range of 2000--6000\,K until melting is observed. \added{In addition, active learning for a 512-atom SiC supercell is conducted at 5000\,K across the same pressure range.} To enhance our data coverage for the decomposition products, additional data is collected for pure carbon (diamond and graphite phases, 2000--4000\,K) and pure silicon (1000--3000\,K) across all pressure conditions.

The final trained MLFF is then used to investigate SiC melting and decomposition under high temperature and pressure conditions through two-phase coexistence simulations, which allowed for the accurate determination of phase transition temperatures.

\paragraph{Large-scale molecular dynamics simulation} The large-scale MD simulations are conducted using GPU-accelerated LAMMPS \cite{thompson2022lammps} with Kokkos parallelization \cite{kokkos} and the FLARE MLFF pairstyle, which has demonstrated exceptional performance in billion-atom catalytic simulations \cite{johansson2022micron}.
During the melting simulations, the NPT ensemble with anisotropic (\textit{aniso}) control of cell dimensions is employed, allowing independent scaling along the $x$, $y$ and $z$ directions through stress components $P_{xx}$, $P_{yy}$, and $P_{zz}$. 
Phase separation exhibits a tendency to minimize the interface between decomposed phases, necessitating anisotropic elongation of the supercell, as illustrated in Fig.~\ref{fig:md_melt}.


\section{Code Availability}

The scripts are available on Github: \url{https://github.com/YuuuXie/SiC_MLMD_phase_diagram}.

\section{Data Availability}

The data and scripts are available on Zenodo: \url{https://doi.org/10.5281/zenodo.14648292}.

\section{Acknowledgement}

We acknowledge Cameron Owen and Anders Johansson for discussions and help with large simulation setup and computational resources. 
We acknowledge Evelyn Hu for helpful discussions and feedback. 
Y.X. is supported by the ``Design \& Assembly of Atomically-Precise Quantum Materials \& Devices'' grant DE-SC0020128 of Department of Energy.
M.W. is supported by National Science Foundation, Office of Advanced Cyberinfrastructure (OAC), under Award No. 2118201.
B.K. and F.S. are supported by the Harvard University Materials Research Science and Engineering Center funded by the National Science Foundation grant DMR-2011754.
The simulation and analysis are done on Harvard Cannon cluster.
This research used resources of the National Energy Research Scientific Computing Center (NERSC), a DOE Office of Science User Facility supported by the Office of Science of the U.S. Department of Energy under Contract No. DE-AC02-05CH11231 using NERSC award BES-ERCAP0024206.

\section{Competing interests}
\noindent
The authors declare no competing financial or non-financial interests.

\section{Author Contributions}

Y.X. initiated the project, performed the training of the ML force field, phase transitions simulations and post analysis.
M.W. contributed to the dataset and code preparation, \added{phase transitions simulations and post analysis}.
S.R. contributed to the DFT settings, and the collection of experimental results.
F.S. guided the analysis of nucleation and decomposition.
B.K. supervised all aspects of the project.
All authors contributed to the writing of the manuscript.

\bibliographystyle{naturemag}
\bibliography{main}

\begin{thebibliography}{10}
\expandafter\ifx\csname url\endcsname\relax
  \def\url#1{\texttt{#1}}\fi
\expandafter\ifx\csname urlprefix\endcsname\relax\def\urlprefix{URL }\fi
\providecommand{\bibinfo}[2]{#2}
\providecommand{\eprint}[2][]{\url{#2}}

\bibitem{madar2004silicon}
\bibinfo{author}{Madar, R.}
\newblock \bibinfo{title}{Silicon carbide in contention}.
\newblock \emph{\bibinfo{journal}{Nature}} \textbf{\bibinfo{volume}{430}}, \bibinfo{pages}{974--975} (\bibinfo{year}{2004}).

\bibitem{madhusudhan2012possible}
\bibinfo{author}{Madhusudhan, N.}, \bibinfo{author}{Lee, K.~K.} \& \bibinfo{author}{Mousis, O.}
\newblock \bibinfo{title}{A possible carbon-rich interior in super-earth 55 cancri e}.
\newblock \emph{\bibinfo{journal}{Astrophys. J. Lett.}} \textbf{\bibinfo{volume}{759}}, \bibinfo{pages}{L40} (\bibinfo{year}{2012}).

\bibitem{katoh2019silicon}
\bibinfo{author}{Katoh, Y.} \& \bibinfo{author}{Snead, L.~L.}
\newblock \bibinfo{title}{Silicon carbide and its composites for nuclear applications--historical overview}.
\newblock \emph{\bibinfo{journal}{Journal of Nuclear Materials}} \textbf{\bibinfo{volume}{526}}, \bibinfo{pages}{151849} (\bibinfo{year}{2019}).

\bibitem{matsunami1997step}
\bibinfo{author}{Matsunami, H.} \& \bibinfo{author}{Kimoto, T.}
\newblock \bibinfo{title}{Step-controlled epitaxial growth of sic: High quality homoepitaxy}.
\newblock \emph{\bibinfo{journal}{Materials Science and Engineering: R: Reports}} \textbf{\bibinfo{volume}{20}}, \bibinfo{pages}{125--166} (\bibinfo{year}{1997}).

\bibitem{yi2007electron}
\bibinfo{author}{Yi, J.}, \bibinfo{author}{He, X.}, \bibinfo{author}{Sun, Y.} \& \bibinfo{author}{Li, Y.}
\newblock \bibinfo{title}{Electron beam-physical vapor deposition of sic/sio2 high emissivity thin film}.
\newblock \emph{\bibinfo{journal}{Applied surface science}} \textbf{\bibinfo{volume}{253}}, \bibinfo{pages}{4361--4366} (\bibinfo{year}{2007}).

\bibitem{zhang2020atomistic}
\bibinfo{author}{Zhang, W.} \& \bibinfo{author}{Van~Duin, A.~C.}
\newblock \bibinfo{title}{Atomistic-scale simulations of the graphene growth on a silicon carbide substrate using thermal decomposition and chemical vapor deposition}.
\newblock \emph{\bibinfo{journal}{Chemistry of Materials}} \textbf{\bibinfo{volume}{32}}, \bibinfo{pages}{8306--8317} (\bibinfo{year}{2020}).

\bibitem{ekimov2004high}
\bibinfo{author}{Ekimov, E.} \emph{et~al.}
\newblock \bibinfo{title}{A high-pressure cell for high-temperature experiments in a toroid-type chamber}.
\newblock \emph{\bibinfo{journal}{Instruments and Experimental Techniques}} \textbf{\bibinfo{volume}{47}}, \bibinfo{pages}{276--278} (\bibinfo{year}{2004}).

\bibitem{togaya1998melting}
\bibinfo{author}{Togaya, M.} \& \bibinfo{author}{Sugiyama, S.}
\newblock \bibinfo{title}{Melting behavior of $\beta$-sic at high pressure}.
\newblock \emph{\bibinfo{journal}{The Review of High Pressure Science and Technology}} \textbf{\bibinfo{volume}{7}}, \bibinfo{pages}{1037--1039} (\bibinfo{year}{1998}).

\bibitem{daviau2017decomposition}
\bibinfo{author}{Daviau, K.} \& \bibinfo{author}{Lee, K.~K.}
\newblock \bibinfo{title}{Decomposition of silicon carbide at high pressures and temperatures}.
\newblock \emph{\bibinfo{journal}{Phys. Rev. B}} \textbf{\bibinfo{volume}{96}}, \bibinfo{pages}{174102} (\bibinfo{year}{2017}).

\bibitem{bhaumik1996modified}
\bibinfo{author}{Bhaumik, S.}, \bibinfo{author}{Divakar, C.}, \bibinfo{author}{Mohan, M.} \& \bibinfo{author}{Singh, A.}
\newblock \bibinfo{title}{A modified high-temperature cell (up to 3300 k) for use with a cubic press}.
\newblock \emph{\bibinfo{journal}{Review of scientific instruments}} \textbf{\bibinfo{volume}{67}}, \bibinfo{pages}{3679--3682} (\bibinfo{year}{1996}).

\bibitem{bhaumik2000synthesis}
\bibinfo{author}{Bhaumik, S.}
\newblock \bibinfo{title}{Synthesis and sintering of monolithic and composite ceramics under high pressures and high temperatures}.
\newblock \emph{\bibinfo{journal}{Metals Materials And Processes}} \textbf{\bibinfo{volume}{12}}, \bibinfo{pages}{215--232} (\bibinfo{year}{2000}).

\bibitem{dolloff1961research}
\bibinfo{author}{Dolloff, R.~T.} \& \bibinfo{author}{Sara, R.}
\newblock \emph{\bibinfo{title}{Research study to determine the phase equilibrium relations of selected metal carbides at high temperatures}}, vol.~\bibinfo{volume}{60} (\bibinfo{publisher}{Aeronautical Systems Division, Air Force Systems Command, US Air Force}, \bibinfo{year}{1961}).

\bibitem{hall1956high}
\bibinfo{author}{Hall, H.~T.}
\newblock \bibinfo{title}{High temperature studies}.
\newblock \emph{\bibinfo{journal}{Bringham Young University: Provo, UT, USA}} \bibinfo{pages}{36} (\bibinfo{year}{1956}).

\bibitem{sokolov2012melting}
\bibinfo{author}{Sokolov, P.~S.}, \bibinfo{author}{Mukhanov, V.~A.}, \bibinfo{author}{Chauveau, T.} \& \bibinfo{author}{Solozhenko, V.~L.}
\newblock \bibinfo{title}{On melting of silicon carbide under pressure}.
\newblock \emph{\bibinfo{journal}{Journal of Superhard Materials}} \textbf{\bibinfo{volume}{34}}, \bibinfo{pages}{339--341} (\bibinfo{year}{2012}).

\bibitem{finocchi1992microscopic}
\bibinfo{author}{Finocchi, F.}, \bibinfo{author}{Galli, G.}, \bibinfo{author}{Parrinello, M.} \& \bibinfo{author}{Bertoni, C.~M.}
\newblock \bibinfo{title}{Microscopic struture of amorphous covalent alloys probed by ab initio molecular dynamics: Sic}.
\newblock \emph{\bibinfo{journal}{Physical review letters}} \textbf{\bibinfo{volume}{68}}, \bibinfo{pages}{3044} (\bibinfo{year}{1992}).

\bibitem{saiz2020ab}
\bibinfo{author}{Saiz, F.}
\newblock \bibinfo{title}{An ab initio study on liquid silicon carbide}.
\newblock \emph{\bibinfo{journal}{J Phys. Chem. Solids}} \textbf{\bibinfo{volume}{137}}, \bibinfo{pages}{109204} (\bibinfo{year}{2020}).

\bibitem{vashishta_interaction_2007}
\bibinfo{author}{Vashishta, P.}, \bibinfo{author}{Kalia, R.~K.}, \bibinfo{author}{Nakano, A.} \& \bibinfo{author}{Rino, J.~P.}
\newblock \bibinfo{title}{Interaction potential for silicon carbide: {A} molecular dynamics study of elastic constants and vibrational density of states for crystalline and amorphous silicon carbide}.
\newblock \emph{\bibinfo{journal}{J. Appl. Phys.}} \textbf{\bibinfo{volume}{101}}, \bibinfo{pages}{103515} (\bibinfo{year}{2007}).

\bibitem{tersoff1989modeling}
\bibinfo{author}{Tersoff, J.}
\newblock \bibinfo{title}{Modeling solid-state chemistry: Interatomic potentials for multicomponent systems}.
\newblock \emph{\bibinfo{journal}{Physical review B}} \textbf{\bibinfo{volume}{39}}, \bibinfo{pages}{5566} (\bibinfo{year}{1989}).

\bibitem{tersoff1994chemical}
\bibinfo{author}{Tersoff, J.}
\newblock \bibinfo{title}{Chemical order in amorphous silicon carbide}.
\newblock \emph{\bibinfo{journal}{Physical Review B}} \textbf{\bibinfo{volume}{49}}, \bibinfo{pages}{16349} (\bibinfo{year}{1994}).

\bibitem{yan2013melting}
\bibinfo{author}{Yan, W.}, \bibinfo{author}{Gao, T.}, \bibinfo{author}{Guo, X.}, \bibinfo{author}{Qin, Y.} \& \bibinfo{author}{Xie, Q.}
\newblock \bibinfo{title}{Melting kinetics of bulk sic using molecular dynamics simulation}.
\newblock \emph{\bibinfo{journal}{Science China Physics, Mechanics and Astronomy}} \textbf{\bibinfo{volume}{56}}, \bibinfo{pages}{1699--1704} (\bibinfo{year}{2013}).

\bibitem{devanathan2007atomistic}
\bibinfo{author}{Devanathan, R.}, \bibinfo{author}{Gao, F.} \& \bibinfo{author}{Weber, W.~J.}
\newblock \bibinfo{title}{Atomistic modeling of amorphous silicon carbide using a bond-order potential}.
\newblock \emph{\bibinfo{journal}{Nucl. Instrum. Methods Phys. Res. B}} \textbf{\bibinfo{volume}{255}}, \bibinfo{pages}{130--135} (\bibinfo{year}{2007}).

\bibitem{kubo2021machine}
\bibinfo{author}{Kubo, A.} \& \bibinfo{author}{Umeno, Y.}
\newblock \bibinfo{title}{Machine-learning-based atomistic model analysis on high-temperature compressive creep properties of amorphous silicon carbide}.
\newblock \emph{\bibinfo{journal}{Materials}} \textbf{\bibinfo{volume}{14}}, \bibinfo{pages}{1597} (\bibinfo{year}{2021}).

\bibitem{macisaac2024genetic}
\bibinfo{author}{MacIsaac, M.}, \bibinfo{author}{Bavdekar, S.}, \bibinfo{author}{Spearot, D.} \& \bibinfo{author}{Subhash, G.}
\newblock \bibinfo{title}{A genetic algorithm trained machine-learned interatomic potential for the silicon--carbon system}.
\newblock \emph{\bibinfo{journal}{The Journal of Physical Chemistry C}} \textbf{\bibinfo{volume}{128}}, \bibinfo{pages}{12213--12226} (\bibinfo{year}{2024}).

\bibitem{liu2024deep}
\bibinfo{author}{Liu, Y.} \emph{et~al.}
\newblock \bibinfo{title}{Deep learning inter-atomic potential for irradiation damage in 3c-sic}.
\newblock \emph{\bibinfo{journal}{Computational Materials Science}} \textbf{\bibinfo{volume}{233}}, \bibinfo{pages}{112693} (\bibinfo{year}{2024}).

\bibitem{ishimaru2003electron}
\bibinfo{author}{Ishimaru, M.}, \bibinfo{author}{Bae, I.-T.} \& \bibinfo{author}{Hirotsu, Y.}
\newblock \bibinfo{title}{Electron-beam-induced amorphization in sic}.
\newblock \emph{\bibinfo{journal}{Physical Review B}} \textbf{\bibinfo{volume}{68}}, \bibinfo{pages}{144102} (\bibinfo{year}{2003}).

\bibitem{vandermause2020fly}
\bibinfo{author}{Vandermause, J.} \emph{et~al.}
\newblock \bibinfo{title}{On-the-fly active learning of interpretable bayesian force fields for atomistic rare events}.
\newblock \emph{\bibinfo{journal}{npj Comput. Mater.}} \textbf{\bibinfo{volume}{6}}, \bibinfo{pages}{1--11} (\bibinfo{year}{2020}).

\bibitem{xie2021bayesian}
\bibinfo{author}{Xie, Y.}, \bibinfo{author}{Vandermause, J.}, \bibinfo{author}{Sun, L.}, \bibinfo{author}{Cepellotti, A.} \& \bibinfo{author}{Kozinsky, B.}
\newblock \bibinfo{title}{Bayesian force fields from active learning for simulation of inter-dimensional transformation of stanene}.
\newblock \emph{\bibinfo{journal}{npj Comput. Mater.}} \textbf{\bibinfo{volume}{7}}, \bibinfo{pages}{1--10} (\bibinfo{year}{2021}).

\bibitem{larsen2016robust}
\bibinfo{author}{Larsen, P.~M.}, \bibinfo{author}{Schmidt, S.} \& \bibinfo{author}{Schi{\o}tz, J.}
\newblock \bibinfo{title}{Robust structural identification via polyhedral template matching}.
\newblock \emph{\bibinfo{journal}{Modelling and Simulation in Materials Science and Engineering}} \textbf{\bibinfo{volume}{24}}, \bibinfo{pages}{055007} (\bibinfo{year}{2016}).

\bibitem{stukowski2009visualization}
\bibinfo{author}{Stukowski, A.}
\newblock \bibinfo{title}{Visualization and analysis of atomistic simulation data with ovito--the open visualization tool}.
\newblock \emph{\bibinfo{journal}{Modelling and simulation in materials science and engineering}} \textbf{\bibinfo{volume}{18}}, \bibinfo{pages}{015012} (\bibinfo{year}{2009}).

\bibitem{gayler1938melting}
\bibinfo{author}{Gayler, M.}
\newblock \bibinfo{title}{Melting point of high-purity silicon}.
\newblock \emph{\bibinfo{journal}{Nature}} \textbf{\bibinfo{volume}{142}}, \bibinfo{pages}{478--478} (\bibinfo{year}{1938}).

\bibitem{dozhdikov2012two}
\bibinfo{author}{Dozhdikov, V.}, \bibinfo{author}{Basharin, A.~Y.} \& \bibinfo{author}{Levashov, P.}
\newblock \bibinfo{title}{Two-phase simulation of the crystalline silicon melting line at pressures from--1 to 3 gpa}.
\newblock \emph{\bibinfo{journal}{The Journal of Chemical Physics}} \textbf{\bibinfo{volume}{137}}, \bibinfo{pages}{054502} (\bibinfo{year}{2012}).

\bibitem{xie2023uncertainty}
\bibinfo{author}{Xie, Y.} \emph{et~al.}
\newblock \bibinfo{title}{Uncertainty-aware molecular dynamics from bayesian active learning for phase transformations and thermal transport in sic}.
\newblock \emph{\bibinfo{journal}{npj Computational Materials}} \textbf{\bibinfo{volume}{9}}, \bibinfo{pages}{36} (\bibinfo{year}{2023}).

\bibitem{krieger1968thermodynamics}
\bibinfo{author}{Krieger, F.~J.}
\newblock \bibinfo{title}{The thermodynamics of the silicon carbide/silicon-carbon vapor system}.
\newblock \bibinfo{type}{Tech. Rep.} (\bibinfo{year}{1968}).

\bibitem{daviau2017zinc}
\bibinfo{author}{Daviau, K.} \& \bibinfo{author}{Lee, K.~K.}
\newblock \bibinfo{title}{Zinc-blende to rocksalt transition in sic in a laser-heated diamond-anvil cell}.
\newblock \emph{\bibinfo{journal}{Physical Review B}} \textbf{\bibinfo{volume}{95}}, \bibinfo{pages}{134108} (\bibinfo{year}{2017}).

\bibitem{kidokoro2017phase}
\bibinfo{author}{Kidokoro, Y.}, \bibinfo{author}{Umemoto, K.}, \bibinfo{author}{Hirose, K.} \& \bibinfo{author}{Ohishi, Y.}
\newblock \bibinfo{title}{Phase transition in sic from zinc-blende to rock-salt structure and implications for carbon-rich extrasolar planets}.
\newblock \emph{\bibinfo{journal}{American Mineralogist: Journal of Earth and Planetary Materials}} \textbf{\bibinfo{volume}{102}}, \bibinfo{pages}{2230--2234} (\bibinfo{year}{2017}).

\bibitem{miozzi2018equation}
\bibinfo{author}{Miozzi, F.} \emph{et~al.}
\newblock \bibinfo{title}{Equation of state of sic at extreme conditions: New insight into the interior of carbon-rich exoplanets}.
\newblock \emph{\bibinfo{journal}{J. Geophys. Res. Planets}} \textbf{\bibinfo{volume}{123}}, \bibinfo{pages}{2295--2309} (\bibinfo{year}{2018}).

\bibitem{sekine1997shock}
\bibinfo{author}{Sekine, T.} \& \bibinfo{author}{Kobayashi, T.}
\newblock \bibinfo{title}{Shock compression of 6h polytype sic to 160 gpa}.
\newblock \emph{\bibinfo{journal}{Phys. Rev. B}} \textbf{\bibinfo{volume}{55}}, \bibinfo{pages}{8034} (\bibinfo{year}{1997}).

\bibitem{tracy2019n}
\bibinfo{author}{Tracy, S.} \emph{et~al.}
\newblock \bibinfo{title}{In situ observation of a phase transition in silicon carbide under shock compression using pulsed x-ray diffraction}.
\newblock \emph{\bibinfo{journal}{Phys. Rev. B}} \textbf{\bibinfo{volume}{99}}, \bibinfo{pages}{214106} (\bibinfo{year}{2019}).

\bibitem{yoshida1993pressure}
\bibinfo{author}{Yoshida, M.}, \bibinfo{author}{Onodera, A.}, \bibinfo{author}{Ueno, M.}, \bibinfo{author}{Takemura, K.} \& \bibinfo{author}{Shimomura, O.}
\newblock \bibinfo{title}{Pressure-induced phase transition in sic}.
\newblock \emph{\bibinfo{journal}{Phys. Rev. B}} \textbf{\bibinfo{volume}{48}}, \bibinfo{pages}{10587} (\bibinfo{year}{1993}).

\bibitem{vandermause2022active}
\bibinfo{author}{Vandermause, J.}, \bibinfo{author}{Xie, Y.}, \bibinfo{author}{Lim, J.~S.}, \bibinfo{author}{Owen, C.~J.} \& \bibinfo{author}{Kozinsky, B.}
\newblock \bibinfo{title}{Active learning of reactive bayesian force fields applied to heterogeneous catalysis dynamics of h/pt}.
\newblock \emph{\bibinfo{journal}{Nature Communications}} \textbf{\bibinfo{volume}{13}}, \bibinfo{pages}{5183} (\bibinfo{year}{2022}).

\bibitem{batzner20223}
\bibinfo{author}{Batzner, S.} \emph{et~al.}
\newblock \bibinfo{title}{E (3)-equivariant graph neural networks for data-efficient and accurate interatomic potentials}.
\newblock \emph{\bibinfo{journal}{Nat. Commun.}} \textbf{\bibinfo{volume}{13}}, \bibinfo{pages}{2453} (\bibinfo{year}{2022}).

\bibitem{musaelian2023learning}
\bibinfo{author}{Musaelian, A.} \emph{et~al.}
\newblock \bibinfo{title}{Learning local equivariant representations for large-scale atomistic dynamics}.
\newblock \emph{\bibinfo{journal}{Nat. Commun.}} \textbf{\bibinfo{volume}{14}}, \bibinfo{pages}{579} (\bibinfo{year}{2023}).

\bibitem{kimoto2016bulk}
\bibinfo{author}{Kimoto, T.}
\newblock \bibinfo{title}{Bulk and epitaxial growth of silicon carbide}.
\newblock \emph{\bibinfo{journal}{Progress in Crystal Growth and Characterization of Materials}} \textbf{\bibinfo{volume}{62}}, \bibinfo{pages}{329--351} (\bibinfo{year}{2016}).

\bibitem{snead1998amorphization}
\bibinfo{author}{Snead, L.}, \bibinfo{author}{Zinkle, S.}, \bibinfo{author}{Hay, J.} \& \bibinfo{author}{Osborne, M.}
\newblock \bibinfo{title}{Amorphization of sic under ion and neutron irradiation}.
\newblock \emph{\bibinfo{journal}{Nuclear Instruments and Methods in Physics Research Section B: Beam Interactions with Materials and Atoms}} \textbf{\bibinfo{volume}{141}}, \bibinfo{pages}{123--132} (\bibinfo{year}{1998}).

\bibitem{zhou2013quantitative}
\bibinfo{author}{Zhou, J.}, \bibinfo{author}{Odqvist, J.}, \bibinfo{author}{Thuvander, M.} \& \bibinfo{author}{Hedstr{\"o}m, P.}
\newblock \bibinfo{title}{Quantitative evaluation of spinodal decomposition in fe-cr by atom probe tomography and radial distribution function analysis}.
\newblock \emph{\bibinfo{journal}{Microscopy and Microanalysis}} \textbf{\bibinfo{volume}{19}}, \bibinfo{pages}{665--675} (\bibinfo{year}{2013}).

\bibitem{sarkar2023nucleation}
\bibinfo{author}{Sarkar, S.~K.}, \bibinfo{author}{Ray, D.}, \bibinfo{author}{Sen, D.} \& \bibinfo{author}{Biswas, A.}
\newblock \bibinfo{title}{Nucleation--growth versus spinodal decomposition in fe--cr alloys: An experimental verification by atom probe tomography and small angle neutron scattering}.
\newblock \emph{\bibinfo{journal}{Microscopy and Microanalysis}} \textbf{\bibinfo{volume}{29}}, \bibinfo{pages}{437--450} (\bibinfo{year}{2023}).

\bibitem{herrero2023cubic}
\bibinfo{author}{Herrero, C.~P.}, \bibinfo{author}{Ramírez, R.} \& \bibinfo{author}{Herrero-Saboya, G.}
\newblock \bibinfo{title}{Cubic silicon carbide under tensile pressure: Spinodal instability}.
\newblock \emph{\bibinfo{journal}{Chemical Physics}} \textbf{\bibinfo{volume}{573}}, \bibinfo{pages}{112005} (\bibinfo{year}{2023}).

\bibitem{maras2016global}
\bibinfo{author}{Maras, E.}, \bibinfo{author}{Trushin, O.}, \bibinfo{author}{Stukowski, A.}, \bibinfo{author}{Ala-Nissila, T.} \& \bibinfo{author}{Jonsson, H.}
\newblock \bibinfo{title}{Global transition path search for dislocation formation in ge on si (001)}.
\newblock \emph{\bibinfo{journal}{Computer Physics Communications}} \textbf{\bibinfo{volume}{205}}, \bibinfo{pages}{13--21} (\bibinfo{year}{2016}).

\bibitem{li2018planar}
\bibinfo{author}{Li, W.}, \bibinfo{author}{Yao, X.} \& \bibinfo{author}{Zhang, X.}
\newblock \bibinfo{title}{Planar impacts on nanocrystalline sic: a comparison of different potentials}.
\newblock \emph{\bibinfo{journal}{Journal of materials science}} \textbf{\bibinfo{volume}{53}}, \bibinfo{pages}{6637--6651} (\bibinfo{year}{2018}).

\bibitem{branicio2018shock}
\bibinfo{author}{Branicio, P.~S.} \emph{et~al.}
\newblock \bibinfo{title}{Shock-induced microstructural response of mono-and nanocrystalline sic ceramics}.
\newblock \emph{\bibinfo{journal}{Journal of Applied Physics}} \textbf{\bibinfo{volume}{123}} (\bibinfo{year}{2018}).

\bibitem{li2017shock}
\bibinfo{author}{Li, W.}, \bibinfo{author}{Yao, X.}, \bibinfo{author}{Branicio, P.}, \bibinfo{author}{Zhang, X.} \& \bibinfo{author}{Zhang, N.}
\newblock \bibinfo{title}{Shock-induced spall in single and nanocrystalline sic}.
\newblock \emph{\bibinfo{journal}{Acta Materialia}} \textbf{\bibinfo{volume}{140}}, \bibinfo{pages}{274--289} (\bibinfo{year}{2017}).

\bibitem{Bartok2015GAP}
\bibinfo{author}{Bartók, A.~P.} \& \bibinfo{author}{Csányi, G.}
\newblock \bibinfo{title}{Gaussian approximation potentials: A brief tutorial introduction}.
\newblock \emph{\bibinfo{journal}{International Journal of Quantum Chemistry}} \textbf{\bibinfo{volume}{115}}, \bibinfo{pages}{1051--1057} (\bibinfo{year}{2015}).

\bibitem{Thompson2015SNAP}
\bibinfo{author}{Thompson, A.}, \bibinfo{author}{Swiler, L.}, \bibinfo{author}{Trott, C.}, \bibinfo{author}{Foiles, S.} \& \bibinfo{author}{Tucker, G.}
\newblock \bibinfo{title}{Spectral neighbor analysis method for automated generation of quantum-accurate interatomic potentials}.
\newblock \emph{\bibinfo{journal}{Journal of Computational Physics}} \textbf{\bibinfo{volume}{285}}, \bibinfo{pages}{316--330} (\bibinfo{year}{2015}).

\bibitem{drautz2019atomic}
\bibinfo{author}{Drautz, R.}
\newblock \bibinfo{title}{Atomic cluster expansion for accurate and transferable interatomic potentials}.
\newblock \emph{\bibinfo{journal}{Phys. Rev. B}} \textbf{\bibinfo{volume}{99}} (\bibinfo{year}{2019}).

\bibitem{thompson2022lammps}
\bibinfo{author}{Thompson, A.~P.} \emph{et~al.}
\newblock \bibinfo{title}{{LAMMPS}-a flexible simulation tool for particle-based materials modeling at the atomic, meso, and continuum scales}.
\newblock \emph{\bibinfo{journal}{Computer Physics Communications}} \textbf{\bibinfo{volume}{271}}, \bibinfo{pages}{108171} (\bibinfo{year}{2022}).

\bibitem{kresse1993ab}
\bibinfo{author}{Kresse, G.} \& \bibinfo{author}{Hafner, J.}
\newblock \bibinfo{title}{Ab initio molecular dynamics for liquid metals}.
\newblock \emph{\bibinfo{journal}{Physical review B}} \textbf{\bibinfo{volume}{47}}, \bibinfo{pages}{558} (\bibinfo{year}{1993}).

\bibitem{perdew1996generalized}
\bibinfo{author}{Perdew, J.~P.}, \bibinfo{author}{Burke, K.} \& \bibinfo{author}{Ernzerhof, M.}
\newblock \bibinfo{title}{Generalized gradient approximation made simple}.
\newblock \emph{\bibinfo{journal}{Physical review letters}} \textbf{\bibinfo{volume}{77}}, \bibinfo{pages}{3865} (\bibinfo{year}{1996}).

\bibitem{ramakers2022effects}
\bibinfo{author}{Ramakers, S.} \emph{et~al.}
\newblock \bibinfo{title}{Effects of thermal, elastic, and surface properties on the stability of sic polytypes}.
\newblock \emph{\bibinfo{journal}{Phys. Rev. B}} \textbf{\bibinfo{volume}{106}}, \bibinfo{pages}{075201} (\bibinfo{year}{2022}).

\bibitem{kokkos}
\bibinfo{author}{Trott, C.~R.} \emph{et~al.}
\newblock \bibinfo{title}{Kokkos 3: Programming model extensions for the exascale era}.
\newblock \emph{\bibinfo{journal}{IEEE Transactions on Parallel and Distributed Systems}} \textbf{\bibinfo{volume}{33}}, \bibinfo{pages}{805--817} (\bibinfo{year}{2022}).

\bibitem{johansson2022micron}
\bibinfo{author}{Johansson, A.} \emph{et~al.}
\newblock \bibinfo{title}{Micron-scale heterogeneous catalysis with bayesian force fields from first principles and active learning}.
\newblock \emph{\bibinfo{journal}{arXiv preprint arXiv:2204.12573}}  (\bibinfo{year}{2022}).

\end{thebibliography}


\begin{thebibliography}{10}
\expandafter\ifx\csname url\endcsname\relax
  \def\url#1{\texttt{#1}}\fi
\expandafter\ifx\csname urlprefix\endcsname\relax\def\urlprefix{URL }\fi
\providecommand{\bibinfo}[2]{#2}
\providecommand{\eprint}[2][]{\url{#2}}

\bibitem{xie2023uncertainty}
\bibinfo{author}{Xie, Y.} \emph{et~al.}
\newblock \bibinfo{title}{Uncertainty-aware molecular dynamics from bayesian active learning for phase transformations and thermal transport in sic}.
\newblock \emph{\bibinfo{journal}{npj Computational Materials}} \textbf{\bibinfo{volume}{9}}, \bibinfo{pages}{36} (\bibinfo{year}{2023}).

\bibitem{birch-murnaghan}
\bibinfo{author}{Birch, F.}
\newblock \bibinfo{title}{Finite elastic strain of cubic crystals}.
\newblock \emph{\bibinfo{journal}{Phys. Rev.}} \textbf{\bibinfo{volume}{71}}, \bibinfo{pages}{809--824} (\bibinfo{year}{1947}).

\bibitem{vinet}
\bibinfo{author}{Vinet, P.}, \bibinfo{author}{Smith, J.~R.}, \bibinfo{author}{Ferrante, J.} \& \bibinfo{author}{Rose, J.~H.}
\newblock \bibinfo{title}{Temperature effects on the universal equation of state of solids}.
\newblock \emph{\bibinfo{journal}{Phys. Rev. B}} \textbf{\bibinfo{volume}{35}}, \bibinfo{pages}{1945--1953} (\bibinfo{year}{1987}).

\bibitem{macisaac2024genetic}
\bibinfo{author}{MacIsaac, M.}, \bibinfo{author}{Bavdekar, S.}, \bibinfo{author}{Spearot, D.} \& \bibinfo{author}{Subhash, G.}
\newblock \bibinfo{title}{A genetic algorithm trained machine-learned interatomic potential for the silicon--carbon system}.
\newblock \emph{\bibinfo{journal}{The Journal of Physical Chemistry C}} \textbf{\bibinfo{volume}{128}}, \bibinfo{pages}{12213--12226} (\bibinfo{year}{2024}).

\bibitem{johansson2022micron}
\bibinfo{author}{Johansson, A.} \emph{et~al.}
\newblock \bibinfo{title}{Micron-scale heterogeneous catalysis with bayesian force fields from first principles and active learning}.
\newblock \emph{\bibinfo{journal}{arXiv preprint arXiv:2204.12573}}  (\bibinfo{year}{2022}).

\bibitem{vashishta_interaction_2007}
\bibinfo{author}{Vashishta, P.}, \bibinfo{author}{Kalia, R.~K.}, \bibinfo{author}{Nakano, A.} \& \bibinfo{author}{Rino, J.~P.}
\newblock \bibinfo{title}{Interaction potential for silicon carbide: {A} molecular dynamics study of elastic constants and vibrational density of states for crystalline and amorphous silicon carbide}.
\newblock \emph{\bibinfo{journal}{J. Appl. Phys.}} \textbf{\bibinfo{volume}{101}}, \bibinfo{pages}{103515} (\bibinfo{year}{2007}).

\bibitem{saiz2020ab}
\bibinfo{author}{Saiz, F.}
\newblock \bibinfo{title}{An ab initio study on liquid silicon carbide}.
\newblock \emph{\bibinfo{journal}{J Phys. Chem. Solids}} \textbf{\bibinfo{volume}{137}}, \bibinfo{pages}{109204} (\bibinfo{year}{2020}).

\bibitem{kubo2021machine}
\bibinfo{author}{Kubo, A.} \& \bibinfo{author}{Umeno, Y.}
\newblock \bibinfo{title}{Machine-learning-based atomistic model analysis on high-temperature compressive creep properties of amorphous silicon carbide}.
\newblock \emph{\bibinfo{journal}{Materials}} \textbf{\bibinfo{volume}{14}}, \bibinfo{pages}{1597} (\bibinfo{year}{2021}).

\bibitem{larsen2016robust}
\bibinfo{author}{Larsen, P.~M.}, \bibinfo{author}{Schmidt, S.} \& \bibinfo{author}{Schi{\o}tz, J.}
\newblock \bibinfo{title}{Robust structural identification via polyhedral template matching}.
\newblock \emph{\bibinfo{journal}{Modelling and Simulation in Materials Science and Engineering}} \textbf{\bibinfo{volume}{24}}, \bibinfo{pages}{055007} (\bibinfo{year}{2016}).

\bibitem{stukowski2009visualization}
\bibinfo{author}{Stukowski, A.}
\newblock \bibinfo{title}{Visualization and analysis of atomistic simulation data with ovito--the open visualization tool}.
\newblock \emph{\bibinfo{journal}{Modelling and simulation in materials science and engineering}} \textbf{\bibinfo{volume}{18}}, \bibinfo{pages}{015012} (\bibinfo{year}{2009}).

\bibitem{zhou2013quantitative}
\bibinfo{author}{Zhou, J.}, \bibinfo{author}{Odqvist, J.}, \bibinfo{author}{Thuvander, M.} \& \bibinfo{author}{Hedstr{\"o}m, P.}
\newblock \bibinfo{title}{Quantitative evaluation of spinodal decomposition in fe-cr by atom probe tomography and radial distribution function analysis}.
\newblock \emph{\bibinfo{journal}{Microscopy and Microanalysis}} \textbf{\bibinfo{volume}{19}}, \bibinfo{pages}{665--675} (\bibinfo{year}{2013}).

\bibitem{sarkar2023nucleation}
\bibinfo{author}{Sarkar, S.~K.}, \bibinfo{author}{Ray, D.}, \bibinfo{author}{Sen, D.} \& \bibinfo{author}{Biswas, A.}
\newblock \bibinfo{title}{Nucleation--growth versus spinodal decomposition in fe--cr alloys: An experimental verification by atom probe tomography and small angle neutron scattering}.
\newblock \emph{\bibinfo{journal}{Microscopy and Microanalysis}} \textbf{\bibinfo{volume}{29}}, \bibinfo{pages}{437--450} (\bibinfo{year}{2023}).

\end{thebibliography}

\end{document}


\large
\date{~\vspace{-5ex}}
\maketitle

\toccontents

\renewcommand{\thefigure}{S\arabic{figure}}
\renewcommand{\thetable}{S\arabic{table}}
\setcounter{figure}{0}
\setcounter{table}{0}

\singlespacing

\section{FLARE parameters}

\begin{table}[H]
    \centering
    \begin{tabular}{ccc}
    \hline
    \hline
        \textbf{FLARE} & \textbf{Parameter} & \textbf{Value} \\
        \hline
        \multirow{4}{10em}{\centering Descriptor} & ACE Descriptor & B2 \\        
        & $n_{max}$ & 8 \\
        & $l_{max}$ & 3 \\
        & Cutoff & 4.0 (\AA) \\
        \hline
        \multirow{7}{10em}{\centering Kernel and GP hyperparameters}  & Kernel & Normalized dot product \\
        & Power ($\xi$) & 2\\
        & Signal variance $\sigma$ & 13.8670\\
        & Energy noise $\sigma_e$ & 0.6039 (eV)\\
        & Force noise $\sigma_f$ & 0.4240 (eV/\AA)\\
        & Stress noise $\sigma_s$ & 0.0063 (eV/\AA$^3$)\\
        & Single atom energies & C: -9.1009 eV, Si: -5.4250 eV\\
        \hline
        \multirow{2}{10em}{\centering Active learning}  & Threshold of calling DFT & 0.03 \\
        & Threshold of adding sparse environments & 0.006 \\
    \hline
    \hline
    \end{tabular}
    \vspace{10pt}
    \caption{Parameters for descriptors, kernels, Gaussian process and active learning in FLARE.}
    \label{tab:FLARE_setup}
\end{table}

\section{DFT parameters}

\begin{table}[H]
    \centering
    \begin{tabular}{cc}
    \hline
    \hline
        \textbf{Parameter} & \textbf{Value} \\
        \hline
       Exchange-correlation functional & PBE \\
       Energy cutoff (ENCUT)  & 800 eV\\
       Smearing (ISMEAR) & Gaussian (0) \\
       Smearing width (SIGMA) & 0.03 eV\\
       Convergence condition (EDIFF) & 1.0e-5 eV for 64 atoms \\
         & 1.0e-4 eV for 512 atoms \\
       K-point  & 3x3x3 for 64 atoms \\
         & 1x1x1 for 512 atoms \\
       \hline
       \hline
    \end{tabular}
    \vspace{10pt}
    \caption{DFT parameters used in VASP.}
    \label{tab:DFT_params}
\end{table}

\newpage
\section{Training data set}

The active learning workflow collected 2088 DFT frames. The offline training from all the DFT data selected 471 frames to add to the final model.

\begin{table}[H]
    \centering
    \begin{tabular}{cccc}
    \hline
    \hline
        \textbf{Data} & \textbf{Temperature (K)} & \textbf{Pressure (GPa)} & \textbf{\# DFT frames}\\
        \hline
        ZB $\leftrightarrow$ RS data \cite{xie2023uncertainty} & 300 - 2000 & 0 - 500 & 680 \\
        Si (start from cubic diamond) & 1000 - 3000 & 0 - 120 & 46 \\
        C (start from cubic diamond) & 2000 - 4000 & 0 - 120 & 55 \\
        C (start from graphite) & 2000 & 0 - 120 & 30 \\
        SiC (64 atoms) & 2000 - 6000 & 0 - 120 & 1128 \\
        SiC (512 atoms) & 5000 & 0 - 120 & 149 \\
    \hline
    \hline
    \end{tabular}
    \vspace{10pt}
    \caption{Training data set. The ZB $\leftrightarrow$ RS data comes from our previous publication \cite{xie2023uncertainty}.}
    \label{tab:training_data}
\end{table}

\newpage
\section{MLFF validation}
\added{\textbf{Bulk Modulus:} In Table.~\ref{tab:bulk_modulus}, we compare the estimation of bulk modulus from the fitting equation of state (EOS) on the electronic energy versus volume (E-V) curve. We consider both Birch–Murnaghan \cite{birch-murnaghan} and Vinet \cite{vinet} EOS forms for fitting. 
To produce the E-V curve, we firstly find the equilibrium lattice constant, and then apply $\pm 2$\% strain and sample 40 volume points. The error between DFT and FLARE estimation is around 0.2--5.4 GPa and the results are close using different EOS for fitting. 
}

\begin{table}[htbp]
\centering
\setlength{\tabcolsep}{8pt}
\begin{tabular}{|c|c|c|c|c|}
\hline
\textbf{Phase} & \multicolumn{2}{c|}{\textbf{FLARE}} & \multicolumn{2}{c|}{\textbf{DFT}} \\
\cline{2-5}
 & \textbf{Birch--Murnaghan (3rd)} & \textbf{Vinet}
 & \textbf{Birch--Murnaghan (3rd)} & \textbf{Vinet} \\
\hline
Zinc Blende & 217.94 & 217.56 & 212.54 & 212.63 \\
Rock Salt   & 265.95 & 265.23 & 265.72 & 265.86 \\
\hline
\end{tabular}
\vspace{5pt}
\caption{Bulk modulus (unit: GPa) from EOS fitting for lattice strain from 0.98 to 1.02.}
\label{tab:bulk_modulus}
\end{table}

\added{\textbf{Lattice constant:} For DFT relaxation, we use VASP with ISIF=3, and apply the following energy and force convergence criteria: EDIFF=1e-8, EDIFFG=-1e-3. For relation using FLARE potential, we use LAMMPS box/relax with conjugate gradient and the same energy and force convergence criteria as above. In Table.~\ref{tab:lattice_constants}, we show that FLARE relaxed lattice constant closely agrees with DFT relaxation.
}

\begin{table}[htbp]
\centering
\setlength{\tabcolsep}{8pt}
\begin{tabular}{|c|c|c|c|}
\hline
\textbf{Phase} & \textbf{Starting structure} & \textbf{FLARE relaxed} & \textbf{DFT relaxed} \\
\hline
Zinc Blende & 4.379 & 4.392 & 4.379 \\
Rock Salt (at 200\,GPa) & 3.800 & 3.574 & 3.586 \\
\hline
\end{tabular}
\vspace{5pt}
\caption{Comparison of lattice constants (unit: \AA) for starting and relaxed structures.}
\label{tab:lattice_constants}
\end{table}

\begin{figure}[htbp]
    \centering
    \includegraphics[width=0.6\textwidth]{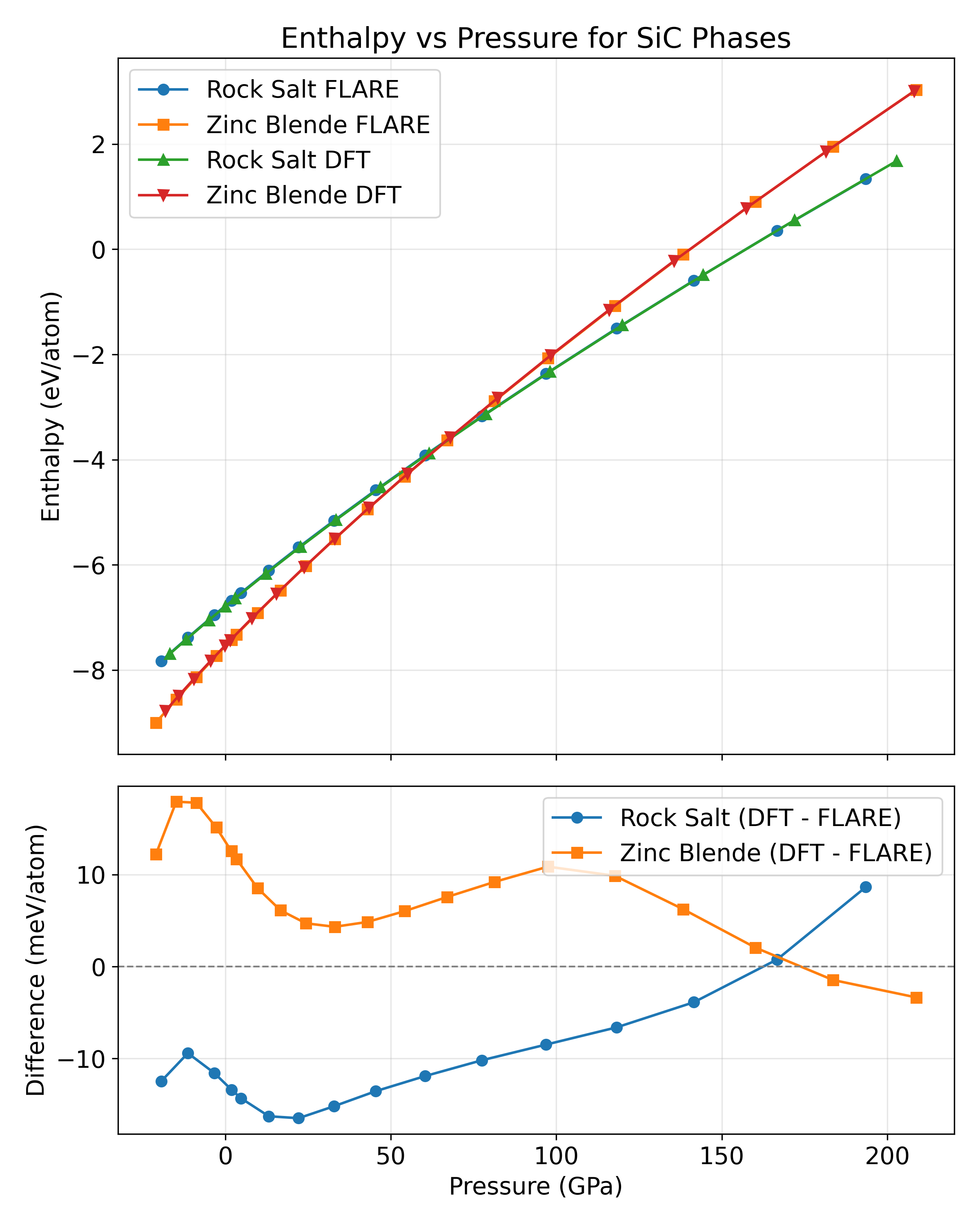}
    \caption{\added{Comparison of FLARE and DFT enthalpies for rock salt and zinc blende phases. Top: Enthalpy per atom from FLARE compared to DFT. Bottom: Difference between DFT and FLARE enthalpy predictions. Note there can be an arbitrary constant shift in enthalpy values, so we apply a single offset to align the curves.}}
    \label{fig:enthalpy_pressure}
\end{figure}

\added{\textbf{Equations of state:} We obtain the enthalpy ($H=U+pV$) by varying the cell volumes ($V$) to determine the electronic energies $U(V)$ and pressure ($p$) predicted by both DFT and the FLARE (our MLFF) for rock salt (B1) and zinc blende (B3) phases. The comparison is demonstrated in Fig.~\ref{fig:enthalpy_pressure}. The errors in the calculated enthalpy are within 15 meV/atom across a pressure range of -20 to 200 GPa. This range fully covers the 0--90 GPa pressure conditions used in our MD simulations. We show that the predicted enthalpy overlaps very well with DFT for both rock salt and zinc blende phases, across the pressure range. In the panel below, we show the per atom difference of enthalpy. Note: Zinc Blende phase is the stable phase at 0 GPa. We follow the validation approach from \cite{macisaac2024genetic} (Fig. 5). The predicted FLARE energies are centered by applying a single offset, calculated from the mean FLARE-DFT energy difference for both phases at their respective electronic energy-volume minima, but we do note that there is arbitrariness in this constant shift.}

\begin{figure}[htbp]
    \centering
    \includegraphics[width=0.5\linewidth]{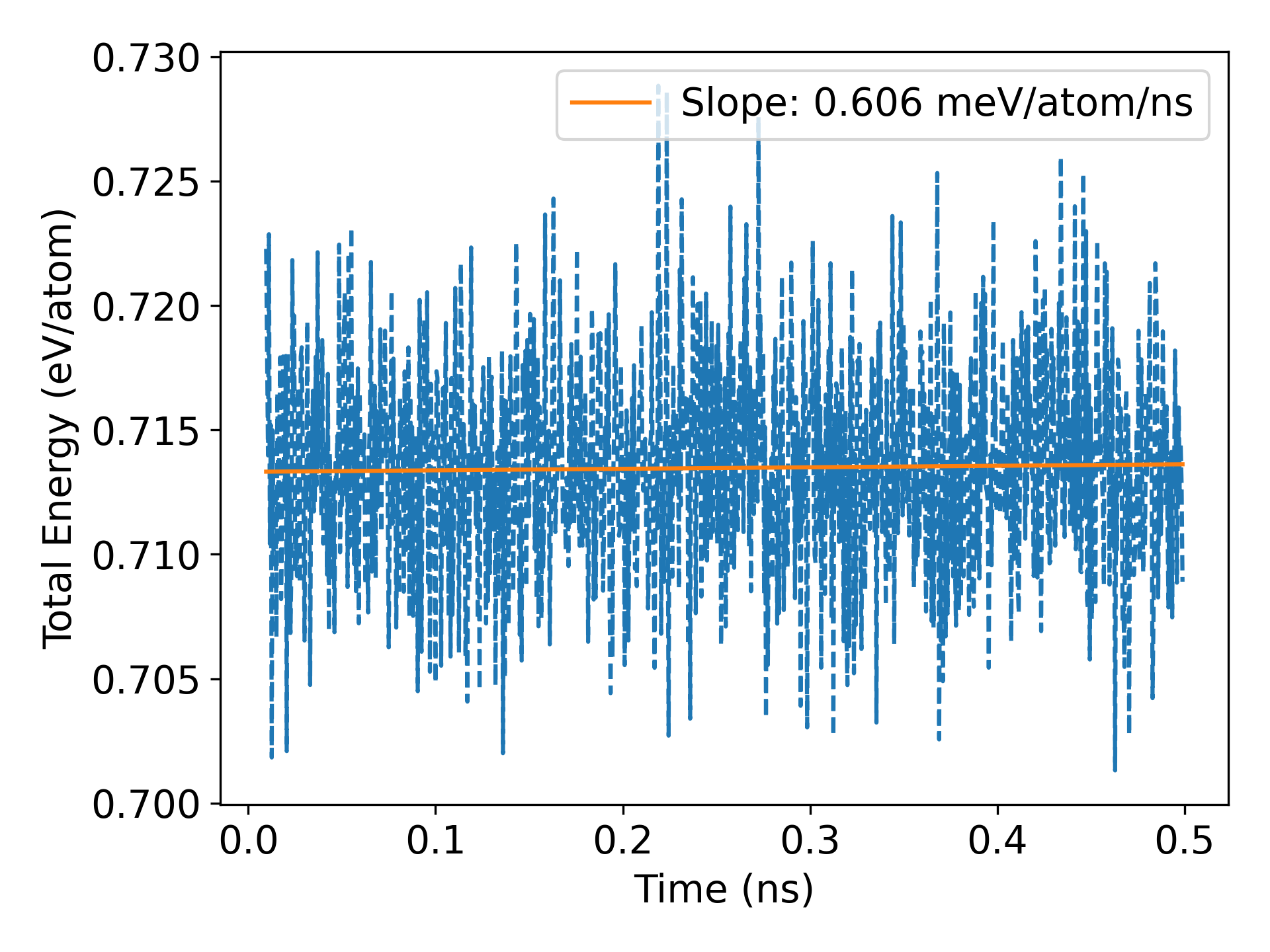}
    \caption{Total energy versus time for a NPT MD simulation at 30 GPa and 3400 K using the FLARE potential.}
    \label{fig:energy_drift}
\end{figure}

\added{\textbf{Energy conservation:} We evaluate the stability of our MD simulations. The energy drift was determined by performing a linear regression on the total energy versus time. This analysis used the initial 0.5 ns of NPT equilibration from the two-phase (B3/B1 $\leftrightarrow$ decomposed Si + C) simulations at various pressures. The average drift decreased significantly with system size: 1.19 $\pm$ 2.62 meV/atom/ns (16,000 atoms), 0.76 $\pm$ 1.50 meV/atom/ns (64,000 atoms), and 0.39 $\pm$ 0.78 meV/atom/ns (128,000 atoms). Fig.~\ref{fig:energy_drift} shows an example NPT simulation (B3 phase at 30 GPa and 3400 K), illustrating that the energy conservation is well conserved in our ML MD simulations.
}

\newpage
\section{MLFF simulation speed}

\added{
Computing speed:
Given one A100 GPU, the performance of MD simulations in LAMMPS is reported in the following table:
}

\begin{table}[htbp]
\centering
\setlength{\tabcolsep}{8pt}
\renewcommand{\arraystretch}{1.2}
\begin{tabular}{|c|c|}
\hline
\textbf{System size} & \textbf{Speed (ns/day)} \\
\hline
16,000 atoms & 5.999 \\
64,000 atoms & 2.965 \\
128,000 atoms & 1.412 \\
\hline
\end{tabular}
\vspace{5pt}
\caption{Simulation performance for systems of different sizes.}
\label{tab:speed}
\end{table}

\added{
We compare the speed of FLARE potential with that of empirical Vashishta potential by running 20,000 steps LAMMPS NPT equilibration MD with 128,000 atoms on a A100 GPU with kokkos, as shown in Table.\ref{tab:speed_flare_vashishta}.
For the speed comparison between FLARE and other ML potentials, we refer to the Figure 7 of Johansson \textit{et al.} \cite{johansson2022micron}, which demonstrates that the speed of FLARE outperforms SNAP and DeePMD by factors of 1.7 and 5.3 at the scale of million-atom simulations. 
}

\begin{table}[htbp]
\centering
\setlength{\tabcolsep}{8pt}
\renewcommand{\arraystretch}{1.2}
\begin{tabular}{|c|c|c|}
\hline
\textbf{Potential} & \textbf{FLARE} & \textbf{Vashishta} \\
\hline
\textbf{Performance} & 1.538 ns/day & 12.095 ns/day \\
\hline
\end{tabular}
\vspace{5pt}
\caption{Comparison of simulation performance between FLARE and Vashishta potentials.}
\label{tab:speed_flare_vashishta}
\end{table}

\newpage
\section{Analysis of decomposed phase}
\subsection{Radial distribution function}

\added{
We note that the radial distribution function (RDF) of the high-temperature amorphous SiC can provide insights into the local atomic arrangements and bonding characteristics of the material, thus demonstrate evidence for congruent vs incongruent melting. 
In particular, if decomposition occurs, we would expect to see dominant peaks of Si-Si and C-C bonds, while a small peak of Si-C bonds, reflecting the presence of separate Si-rich and C-rich regions. 
Conversely, in a congruently melted state, the RDF would show a dominant Si-C peak, with smaller peaks for Si-Si and C-C bonds, indicating a more homogeneous mixture of Si and C atoms.
}

\begin{figure}[H]
    \centering
    \includegraphics[width=\textwidth]{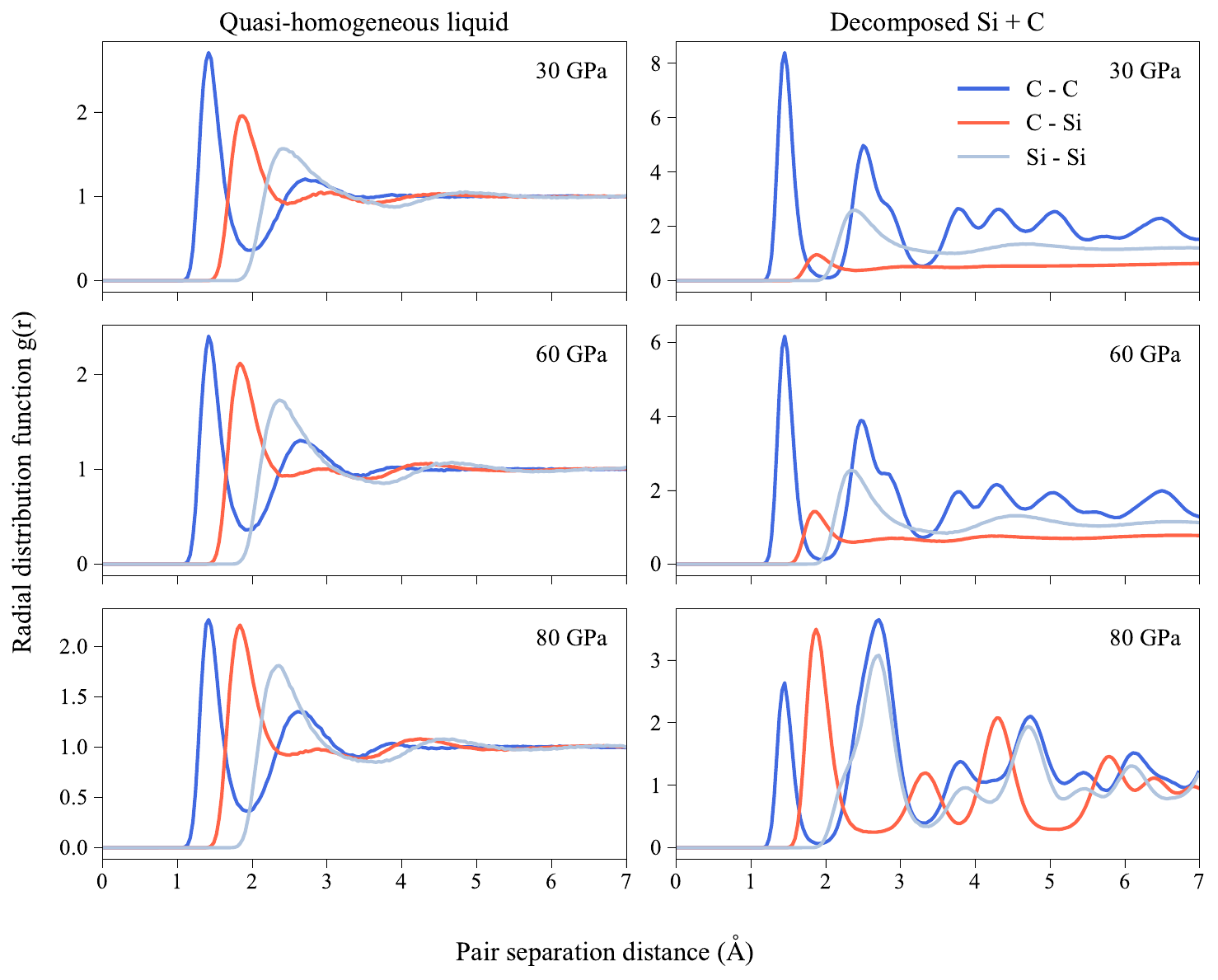}
    \caption{Radial distribution function of the high-temperature amorphous SiC.}
    \label{sub-fig:rdf}
\end{figure}

\added{
In Fig.~\ref{sub-fig:rdf}, we present the RDF of the SiC generated by our FLARE model large-scale MD simulations. At homogeneous liquid phase (left column) the RDF shows a dominant Si-C peak, with smaller peaks for Si-Si and C-C bonds. At decomposed phase (right column), the RDF shows dominant peaks of Si-Si and C-C bonds.
}

\begin{figure}[H]
    \centering
    \includegraphics[width=\textwidth]{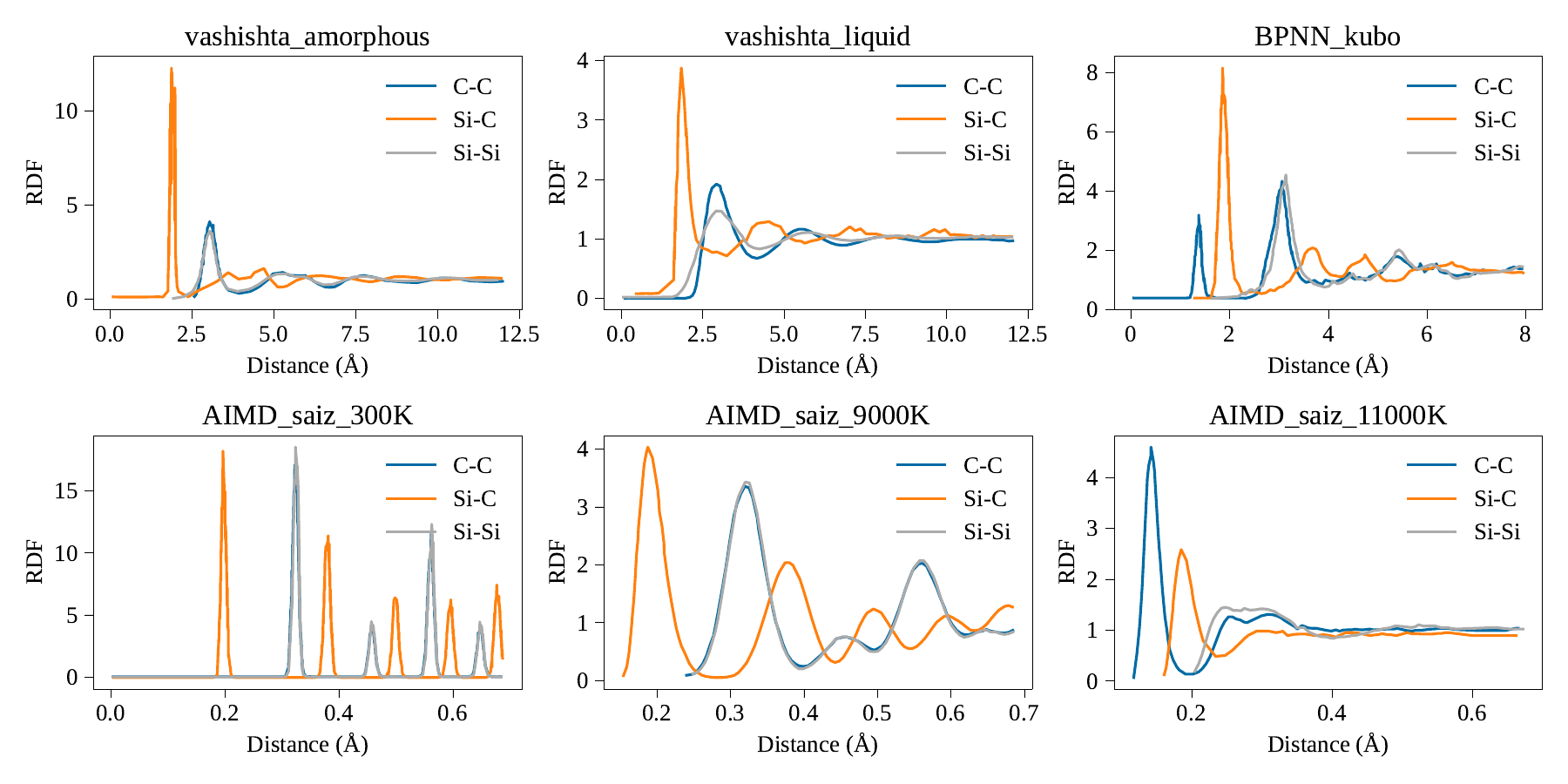}
    \caption{Radial distribution functions of C-C, Si-C, and Si-Si, from Vashishta empirical potential \cite{vashishta_interaction_2007}, \textit{ab initio} MD \cite{saiz2020ab} (AIMD), and Behler-Parrinello neural network \cite{kubo2021machine} (BPNN).}
    \label{sub-fig:rdf_comparison}
\end{figure}

\added{
The RDF of the high-temperature amorphous SiC generated by Vashishta empirical potential \cite{vashishta_interaction_2007}, \textit{ab initio} MD \cite{saiz2020ab} (AIMD), and Behler-Parrinello neural network \cite{kubo2021machine} (BPNN) are compared in Fig.~\ref{sub-fig:rdf_comparison}. In the simulations with Vashishta potential, the Si-C peak dominates the RDF, while the Si-Si and C-C peaks are much smaller. This indicates that the amorphous and liquid SiC from Vashishta potential is more like a homogeneous mixture of Si and C atoms, suggesting congruent melting.
The BPNN model improves the situation, where the C-C peak at 1.5\,\AA\ appears, but the Si-C peak is still noticeably higher than the other two peaks, indicating the formation of small carbon clusters but not full decomposition.
While in AIMD simulations, even when the temperature raises to 9000 K, the RDF peaks are still similar to 300 K but broadened. However, when the temperature is increased to 11000 K, the C-C peak begins to dominate, indicating the formation of larger carbon clusters. Despite showing strong evidence and trend, AIMD can not reach the full decomposition and quantitatively capture the transition temperature and pressure because of the limited system size and time scale.
}

\newpage
\subsection{Strutural analysis}

\begin{figure}[h]
    \centering
    \includegraphics[width=\textwidth]{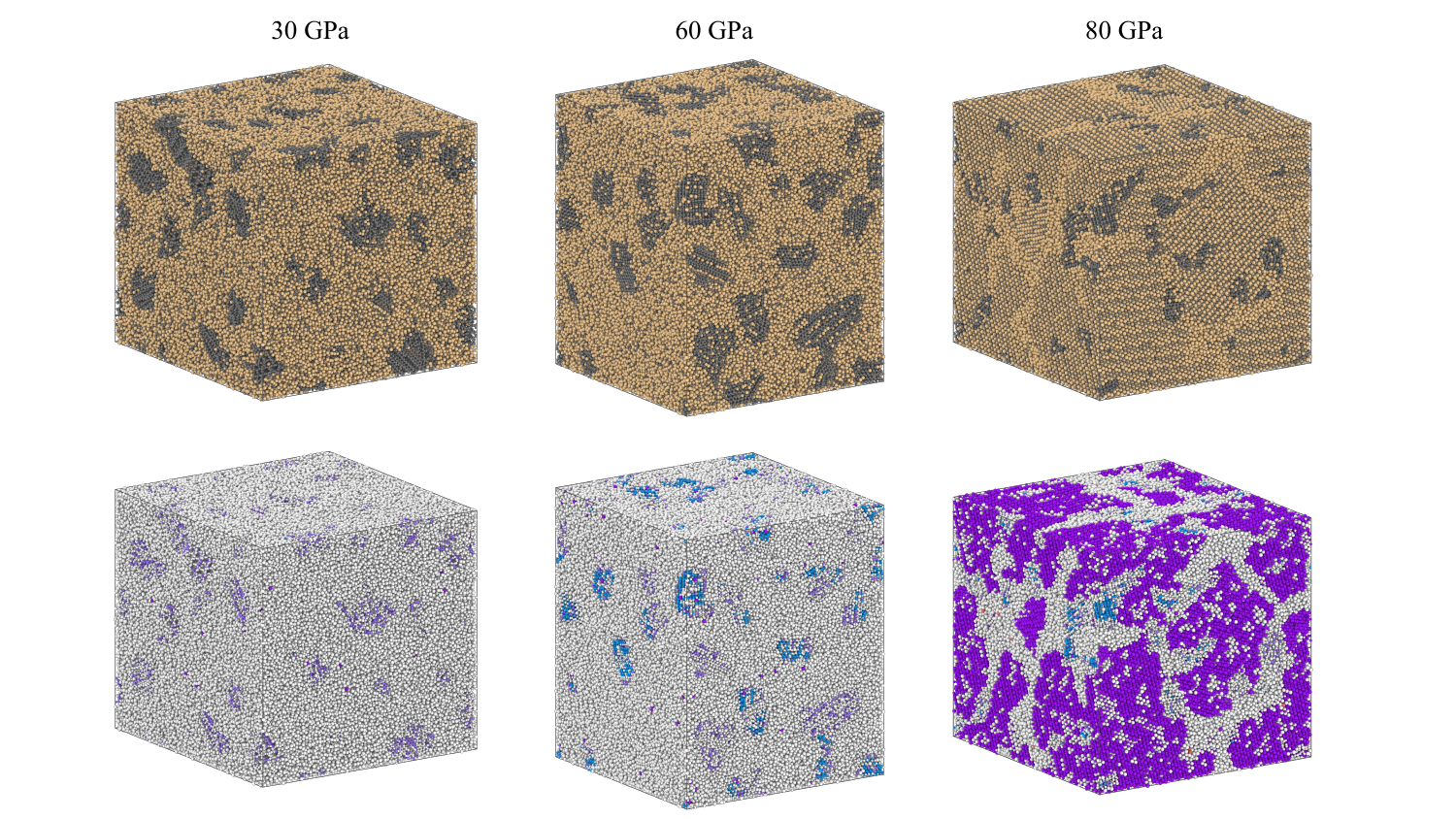}
    \caption{0.5M atoms configuration at 30 GPa, 60 GPa and 80 GPa. Top row: the decomposed configuration at the end of cooling process at 3000 K with atoms colored by chemical elements (Orange: Si, black: C). Bottom row: the same configuration as the top row, but colored by crystal types classified by PTM \added{with RMSD of 0.16} (Blue: cubic diamond, light purple: graphene, dark purple: simple cubic).}
    \label{fig:decomposition_snapshots}
\end{figure}

\added{To analyze the structures formed in the MD simulations, we employed the following methods. For cooling simulations that eventually reach decomposition, we used polyhedral template matching (PTM) \cite{larsen2016robust} to identify simple cubic, cubic diamond, and graphene structures as shown in Fig.~\ref{fig:decomposition_snapshots} and Fig.~\ref{fig:msd_nanocluster}. For B1 $\leftrightarrow$ liquid two-phase simulations in Fig.~\ref{fig:two_phase_B1}, we applied PTM with root-mean-square deviation (RMSD) of 0.2 to identify simple cubic structures and determine the fraction of B1 SiC. For B3 $\leftrightarrow$ decomposed Si + C two-phase simulations in Fig.~\ref{fig:two_phase_B3}, we used ``identify diamond structure" method implemented in OVITO \cite{stukowski2009visualization} to detect cubic diamond SiC structures. However, to resolve the potential ambiguity arising from both B3-SiC and carbon diamond in the decomposed phase being identified as cubic diamond structures, we imposed an additional constraint: atoms were classified as B3 SiC only if they exhibited cubic diamond structure and their carbon atoms were not located within any carbon nanoclusters identified through cluster analysis.}


\added{To distinguish carbon diamonds in the decomposed phase from B3-SiC, we calculated the fraction of cubic diamond-structured carbon atoms located within large carbon nanoclusters ($>100$ atoms, excluding small transient clusters). A fraction near 1 confrims that these diamond carbons belong to the decomposed phase rather than B3-SiC. The analysis showed fractions of 1.0000 (at 30 GPa), 0.9995 (at 60 GPa), and 0.9999 (at 80 GPa). For context, the total percentage of identified diamond carbons (relative to all C atoms) at these pressures was 0.0031\%, 8.6\%, and 5.4\%, respectively. } 

\newpage
\subsection{Mean squared displacement analysis}

\begin{figure}[h]
    \centering
    \includegraphics[width=\textwidth]{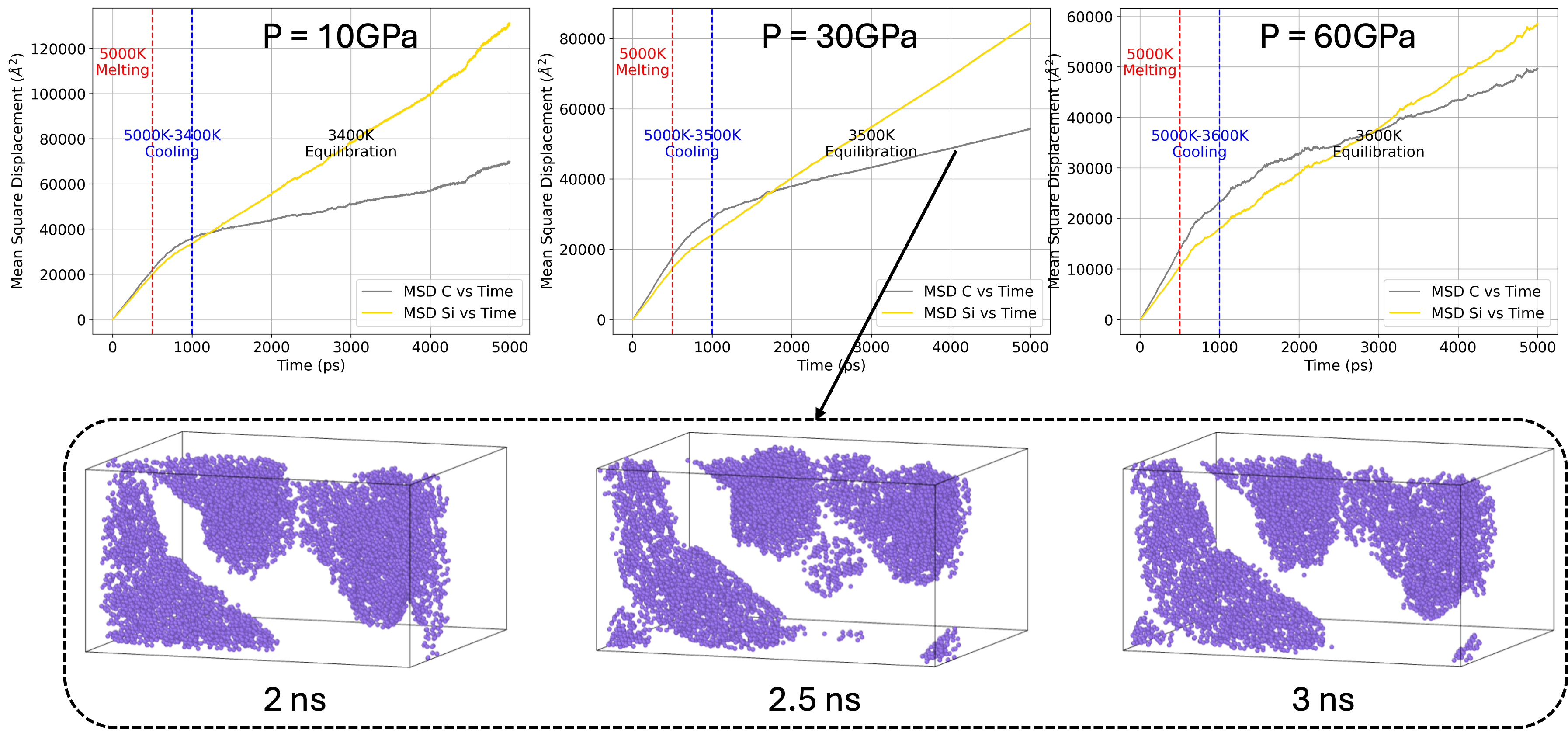}
    \caption{\added{Mean squared displacement (MSD) of C and Si and evolution of carbon nanoclusters at 30 GPa. Top row: MSD of C (grey) and Si (gold) atoms from 64,000-atoms MD simulations, shown at 10 GPa, 30 GPa and 60 GPa. Each simulation involves three stages: (1) a 500 ps melting run at 5000 K, (2) a 500 ps cooling run toward a lower decomposition temperature, and (3) a 4 ns equilibration run at the final temperatures of 3400 K (10 GPa), 3500 K (30 GPa), and 3600 K (60 GPa). The MSD calculation uses the first frame as the reference point, demonstrating the decomposition of SiC into liquid Si and solid C. Bottom row: snapshots illustrating the evolution of carbon nanoclusters during equilibration from the 30 GPa simulation. Carbon atoms are identified as a graphene-like structure (light purple) using PTM with RMSD cutoff of 0.2. The images show the formation and growth of the carbon nanocluster between 2 ns and 3 ns of the equilibration run.}}
    \label{fig:msd_nanocluster}
\end{figure}

\added{Fig.~\ref{fig:msd_nanocluster} illustrates the decomposition of SiC into solid carbon structure and liquid silicon. In the top row, the initial steep, linear increase in the MSD for both C (grey) and Si (gold) during the 5000 K stage confirms the liquid state of both elements. As the system cools to the final equilibration temperatures (3400 K to 3600 K), the distinct changes in slope imply the onset of incongruent melting (decomposition):}

\begin{itemize}
    \item The MSD of Si maintains a significantly steep, linear slope, confirming its transition to a stable liquid phase.
    \item The MSD of C exhibits a near-plateauing trend, indicating a rapid decrease in atomic motion consistent with the formation of a solid phase.
\end{itemize}

\added{The fact that the carbon MSD does not completely flatten is due to the internal movement and rearrangement of the solid carbon clusters. The bottom row confirms this interpretation, visually demonstrating the solid phase formation at 30 GPa. Carbon atoms identified as a graphene-like structure (light purple) nucleates and the solid carbon nanoclusters move and evolve during the equilibration stage.}

\newpage
\subsection{Volume change during decomposition MD}

\begin{figure}[h]
    \centering
    \includegraphics[width=0.9\textwidth]{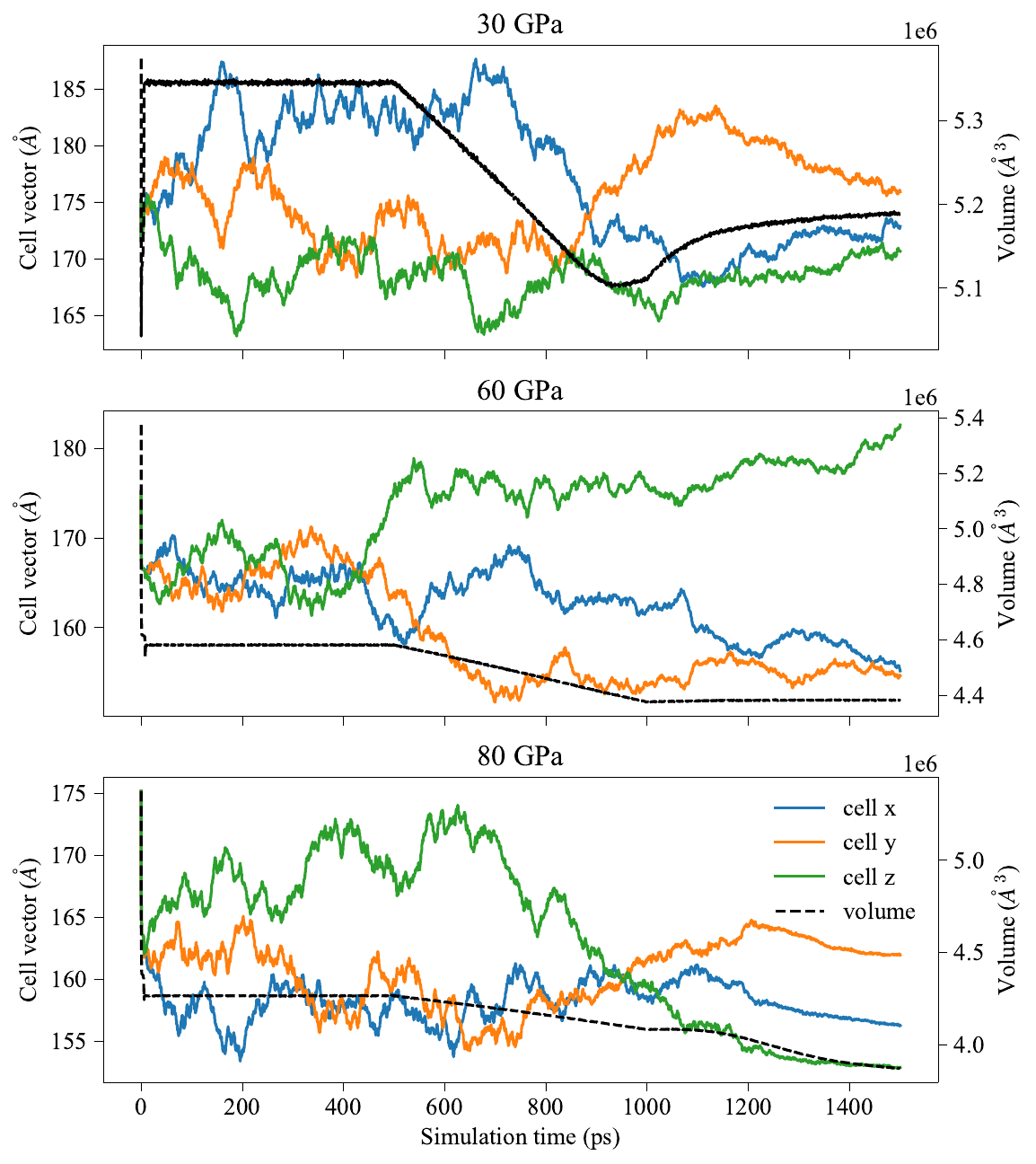}
    \caption{0.5M atoms cell dimension and volume at 30 GPa, 60 GPa and 80 GPa.}
    \label{fig:cell_volume_decompose_MD}
\end{figure}

\newpage
\subsection{Spatial correlation function}

To investigate the spatial distribution of concentration of the decomposed structure, one approach is to compute the radial distribution function.
However,  the characteristic size of the clusters is a few nanometers. When looking at individual particles, the calculation of radial distribution function with cutoff beyond 2 nm can be very slow.

Therefore, we instead look at the continuous model with local concentration. 
Basically, we discretize the super cell into small bins and compute the local carbon concentration which is the number of carbon atoms in the bin divided by the volume of the bin. 
Then we compute the spatial correlation function w.r.t. distance $g_{AB}(r) = \langle c_A(0) c_B(r) \rangle$ with $A$ and $B$ representing chemical elements and $c(r)$ is the concentration at distance $r$.

In the C-C correlation function $g_{CC}(r)$ as shown in the figure below, we observe the peak at around 4 nm corresponding to the statistically averaged distance between two C-rich regions. It also indicates the wavelength of the spinodal decomposition mode of SiC phase separation \cite{zhou2013quantitative,sarkar2023nucleation}. 

\begin{figure}[h]
    \centering
    \includegraphics[width=0.7\textwidth]{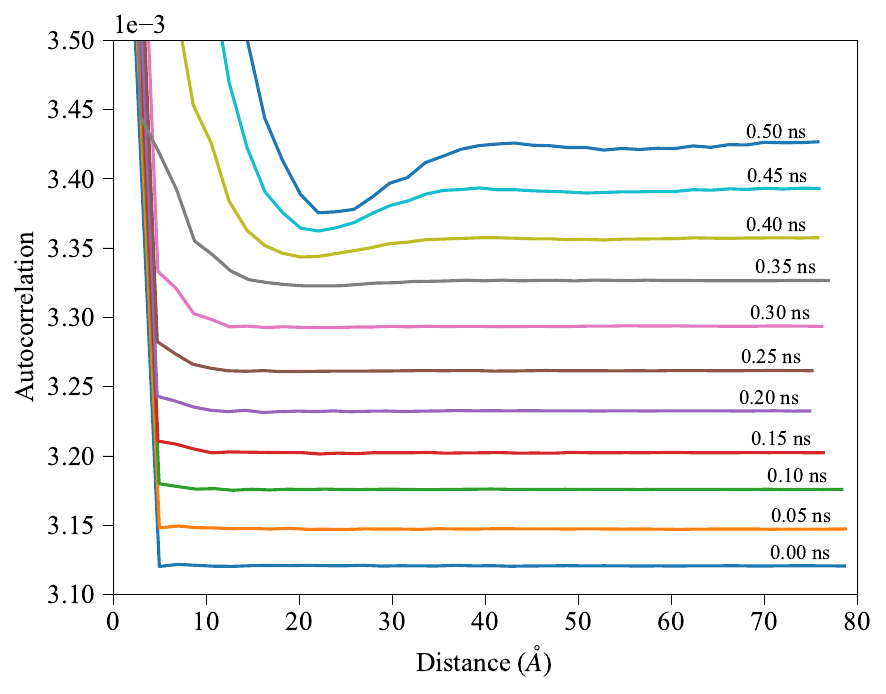}
    \caption{Spatial correlation function of C-C. P=60 GPa, cooling down from 5000K (0 ns) to 3000K (0.5 ns).}
    \label{fig:C-C_autocorr}
\end{figure}

\newpage
\section{Two-phase coexistence simulations}
\subsection{High temperature transition from homogeneous liquid to decomposed Si + C} \label{sec:two_phase_liquid}
\begin{figure}[h]
    \centering
    \includegraphics[width=0.7\textwidth]{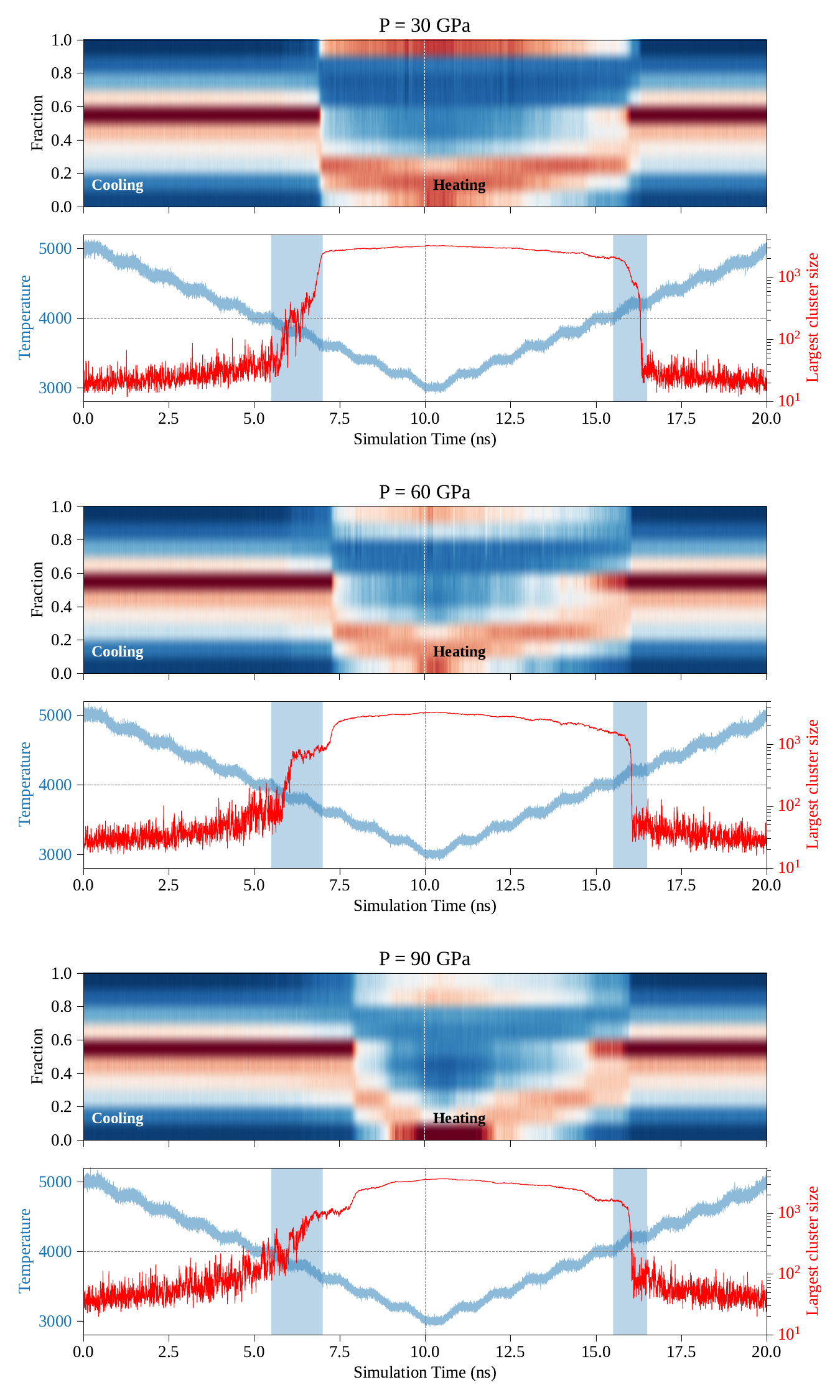}
    \caption{Cooling and heating of SiC for the transition between Si + C and near homogeneous liquid at 60 and 90 GPa.}
    \label{fig:local_concentration}
\end{figure}

\begin{figure}[H]
    \centering
    \includegraphics[width=0.8\textwidth]{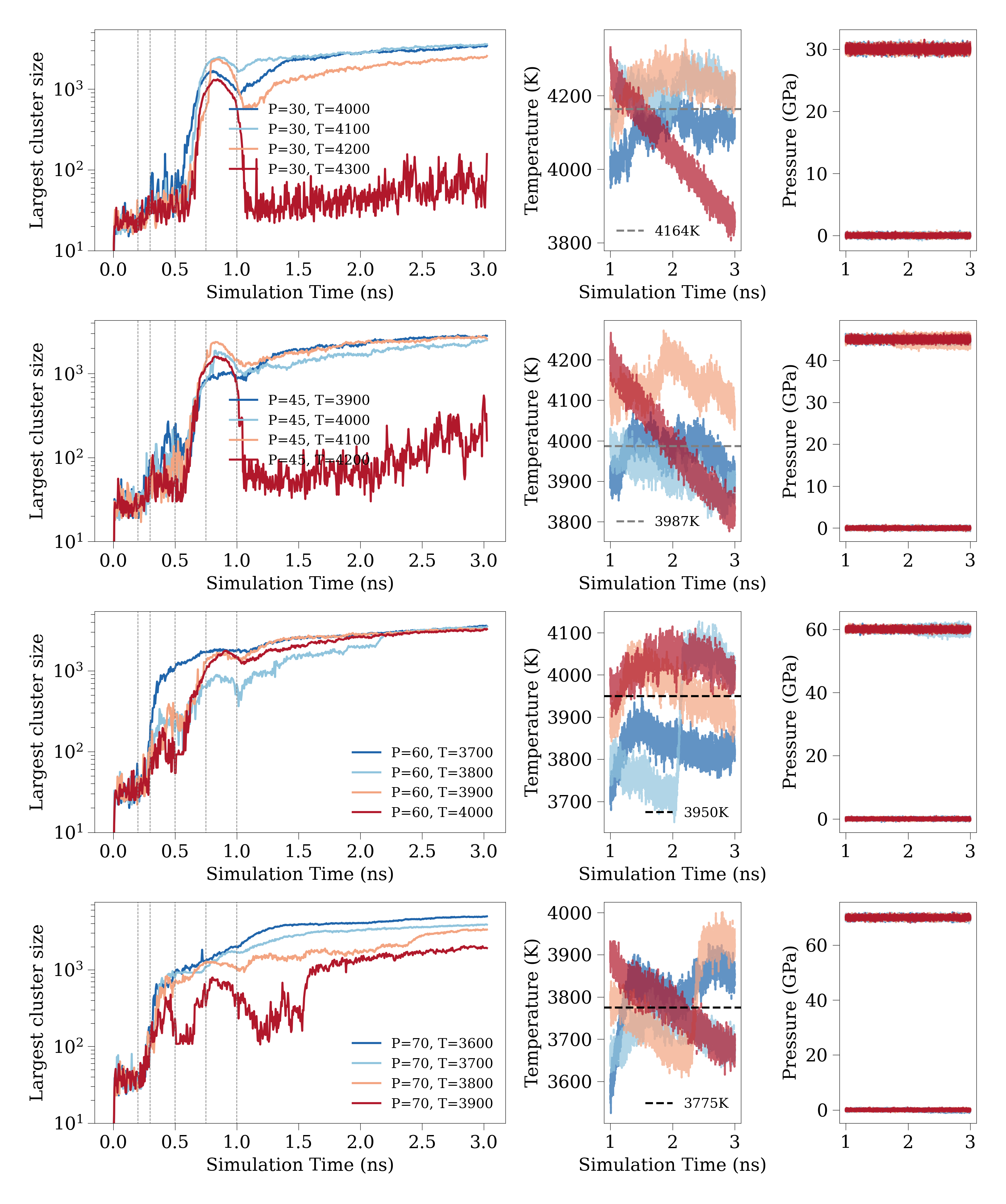}
    \caption{\added{Two-phase coexistence MD anlysis at 30, 45, 60, and 70 GPa. Panels from left to right tracks time evolution of: the size of largest carbon cluster, the system temperature, and the full pressure tensor components, including both the diagonal ($P_{xx},P_{yy},P_{zz}$) and off-diagonal ($P_{xy},P_{xz},P_{yz}$).}}
    \label{fig:liquid_two_phase}
\end{figure}

\added{To compute the averaged transition temperature, we only include runs with the average largest cluster size exceed 100 atoms during the last 250 ps. For example, at 30 GPa, the trajectory at 4300 K is excluded from the average since the carbon clusters diminish and the temperature does not converge. More detailed criteria are mentioned in Supplementary section~\ref{sec:two_phase_B3}.}

\newpage
\subsection{Snapshots of two-phase coexistence simulation}
\begin{figure}[H]
    \centering
    \includegraphics[width=\textwidth]{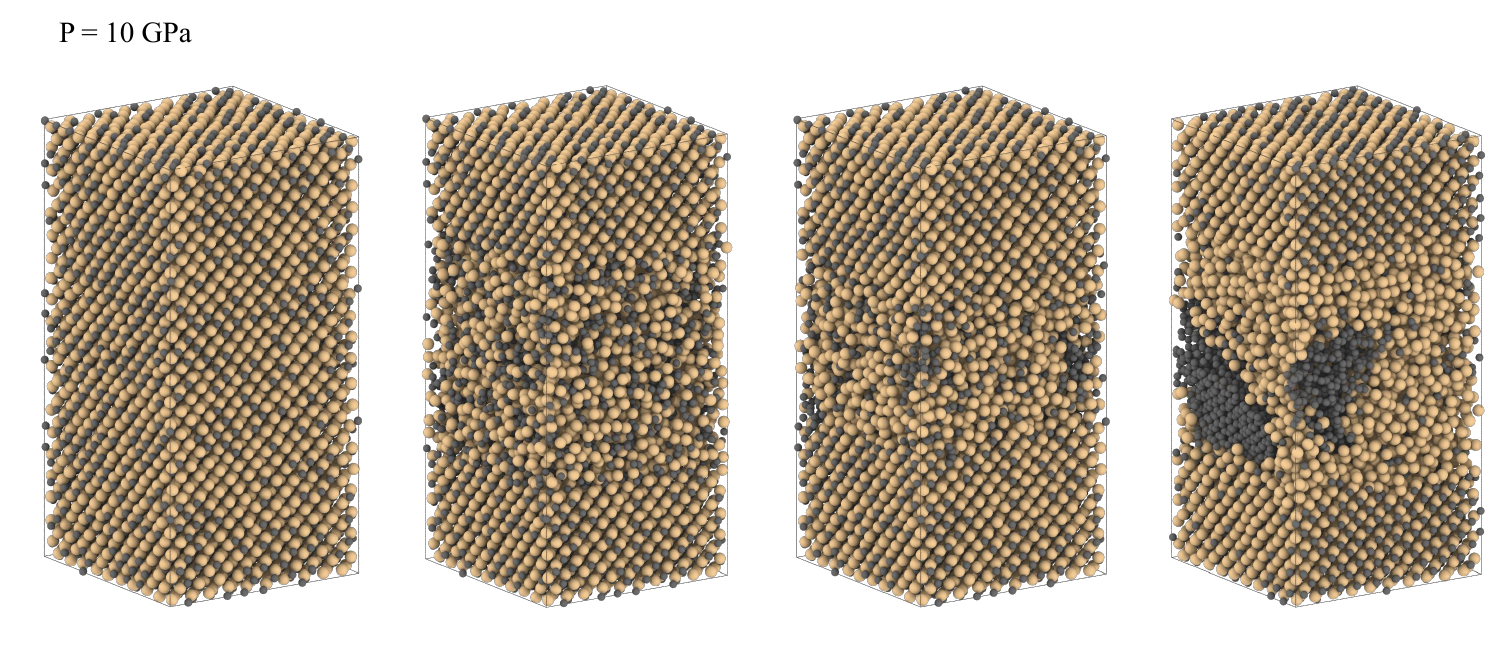}
    \includegraphics[width=\textwidth]{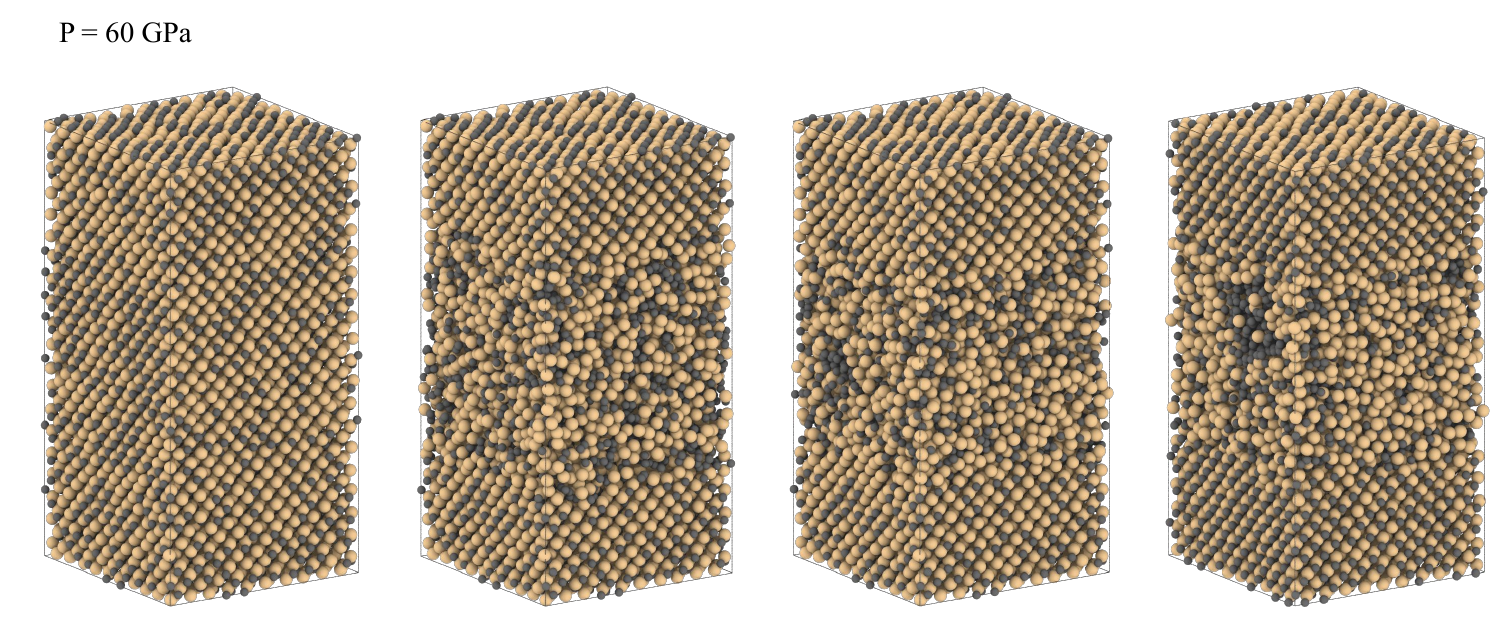}
    \caption{Two-phase MD snapshots at pressures 10, 30 and 60 GPa, starting at B3 phase (zinc blende, cubic diamond lattice).}
    \label{fig:two_phase_snapshot}
\end{figure}

\newpage
\subsection{Low temperature transition from B3 to decomposed Si + C}\label{sec:two_phase_B3}
\begin{figure}[H]
    \centering
    \includegraphics[width=0.85\textwidth]{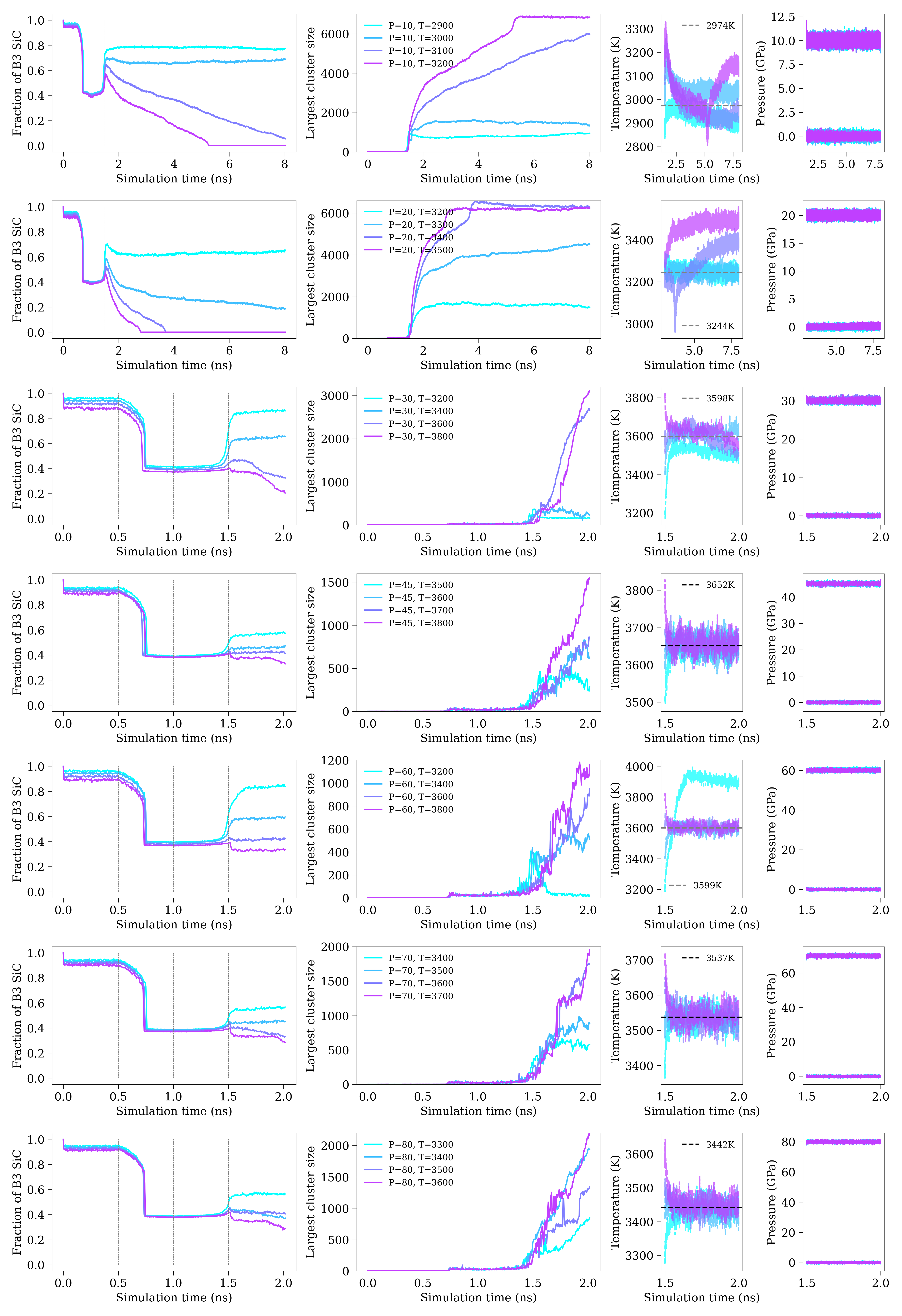}
    \caption{\added{Two-phase MD structural and cluster analysis for 10, 20, 30, 45, 60, 70 and 80 GPa. Panels from left to right tracks time evolution of: the fraction of B3 SiC, the size of largest carbon cluster, the system temperature, and the full pressure tensor components, including both the diagonal ($P_{xx},P_{yy},P_{zz}$) and off-diagonal ($P_{xy},P_{xz},P_{yz}$).}}
    \label{fig:two_phase_B3}
\end{figure}

\added{
We analyzed the coexistence by calculating the fraction of B3-SiC, defined as the number of atoms in cubic diamond structures divided by the total number of atoms. We used OVITO's ``identify diamond structure" modifier to detect all cubic diamond configurations, then excluded atoms belonging to large carbon clusters (identified through cluster analysis). Therefore, the remaining diamond structures represent exclusively the B3-SiC phase.}

\added{To determine the transition temperatures, we average the temperature during the final $NPH$ coexistence simulations that meet specific criteria. We only include runs where:}

\begin{itemize}
\item \added{the average B3/B1 solid fraction remain between 0.15 and 0.85, and}
\item \added{the average largest cluster size exceed 100 atoms during the last 250 ps.}
\end{itemize}

\added{These criteria confirm the stable coexistence of both B3/B1 and decomposed Si + C phases. Runs failing these criteria, such as simulations at 10 GPa with initial estimated temperatures ($T_{es}$) of 3100 K and 3200 K (resulting in complete decomposition, with no B3 phase), or the 60 GPa run at 3200 K (resulting in full B3 solidification, with no decomposed Si+C phase), are excluded. In the third column of Fig.~\ref{fig:two_phase_B3}, a solid line indicates that all runs at a given pressure are included, while a dashed line indicates that some runs are excluded from the transition temperature calculation based on the criteria mentioned above. This convention applies to all other two-phase simulation plots. Finally, we use the later half of the $NPH$ trajectory to compute the final transition temperature. }

\added{At 10 GPa and 20 GPa, the $NPH$ simulations required longer run times and a swtich from an isotropic to an anisotropic barostat setting to achieve pressure convergence. This slow convergence is likely related to the more substantial changes in lateral pressures ($P_{xx}, P_{yy}$) that occurred during the interface creation stages (stages 2 and 3) compared to the higher-pressure simulations (e.g., above 30 GPa).}

\newpage
\subsection{Transition from B1 to liquid SiC phase}
\begin{figure}[H]
    \centering
    \includegraphics[width=\textwidth]{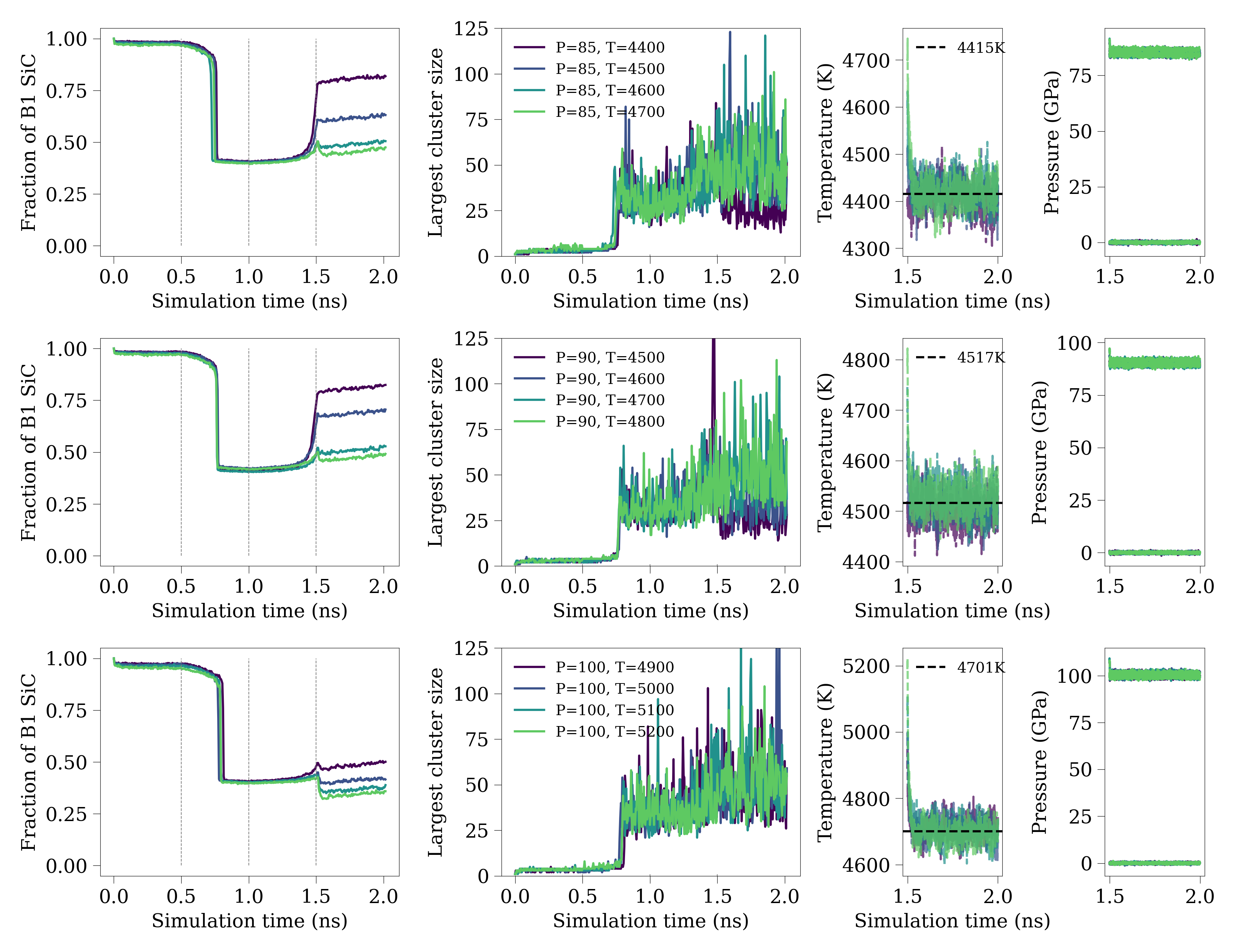}
    \caption{\added{Two-phase MD structural and cluster analysis for 85, 90, 100 GPa. Panels from left to right tracks time evolution of: the fraction of B1 SiC, the size of largest carbon cluster, the system temperature, and the full pressure tensor components, including both the diagonal ($P_{xx},P_{yy},P_{zz}$) and off-diagonal ($P_{xy},P_{xz},P_{yz}$).}}
    \label{fig:two_phase_B1}
\end{figure}

\newpage
\subsection{Influence of defect concentration}
\begin{figure}[H]
    \centering
    \includegraphics[width=\textwidth]{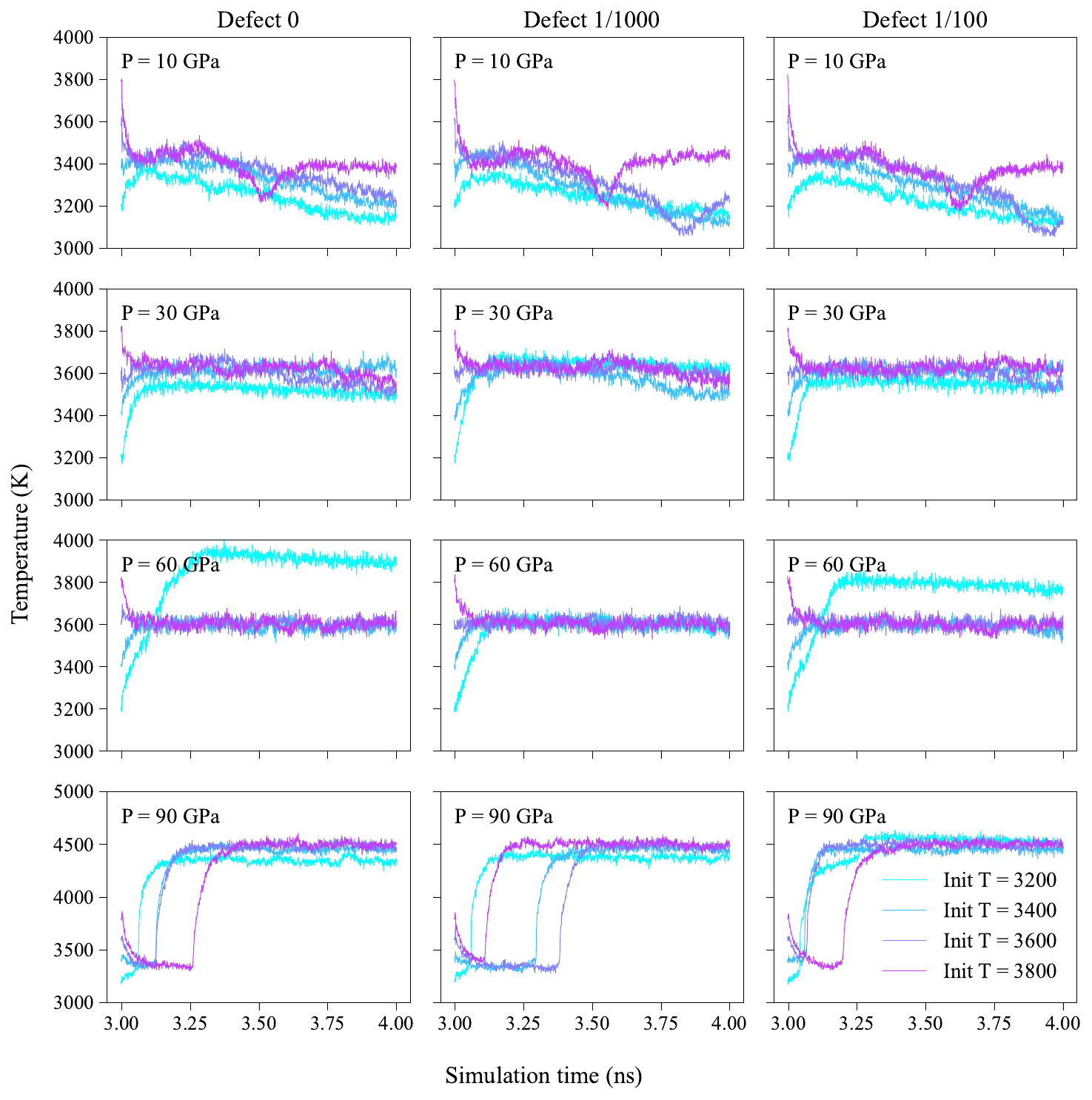}
    \caption{Two-phase results with different defect concentrations.}
    \label{fig:defect_effect}
\end{figure}

\newpage
\subsection{Effect of simulation size on phase transition temperature}
\begin{figure}[htbp]
    \centering
    \includegraphics[width=\textwidth]{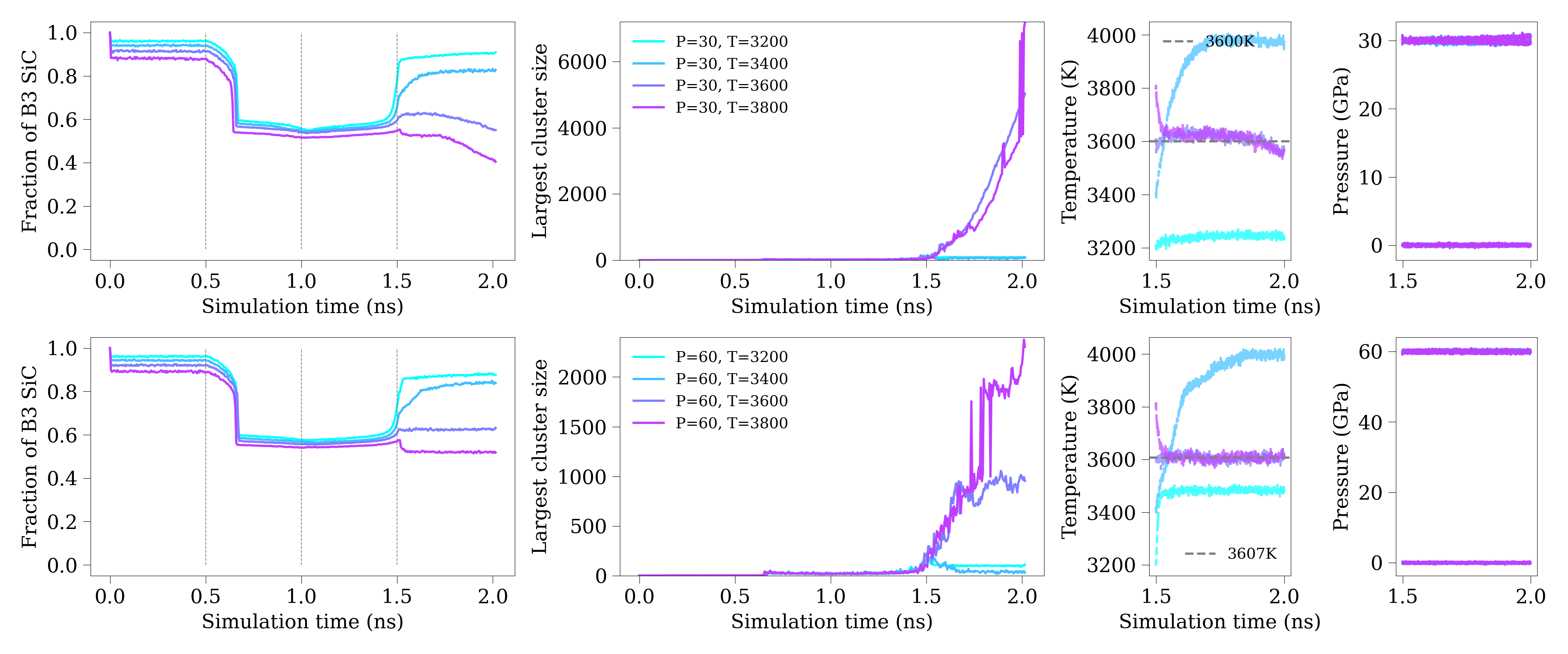}
    \includegraphics[width=\textwidth]{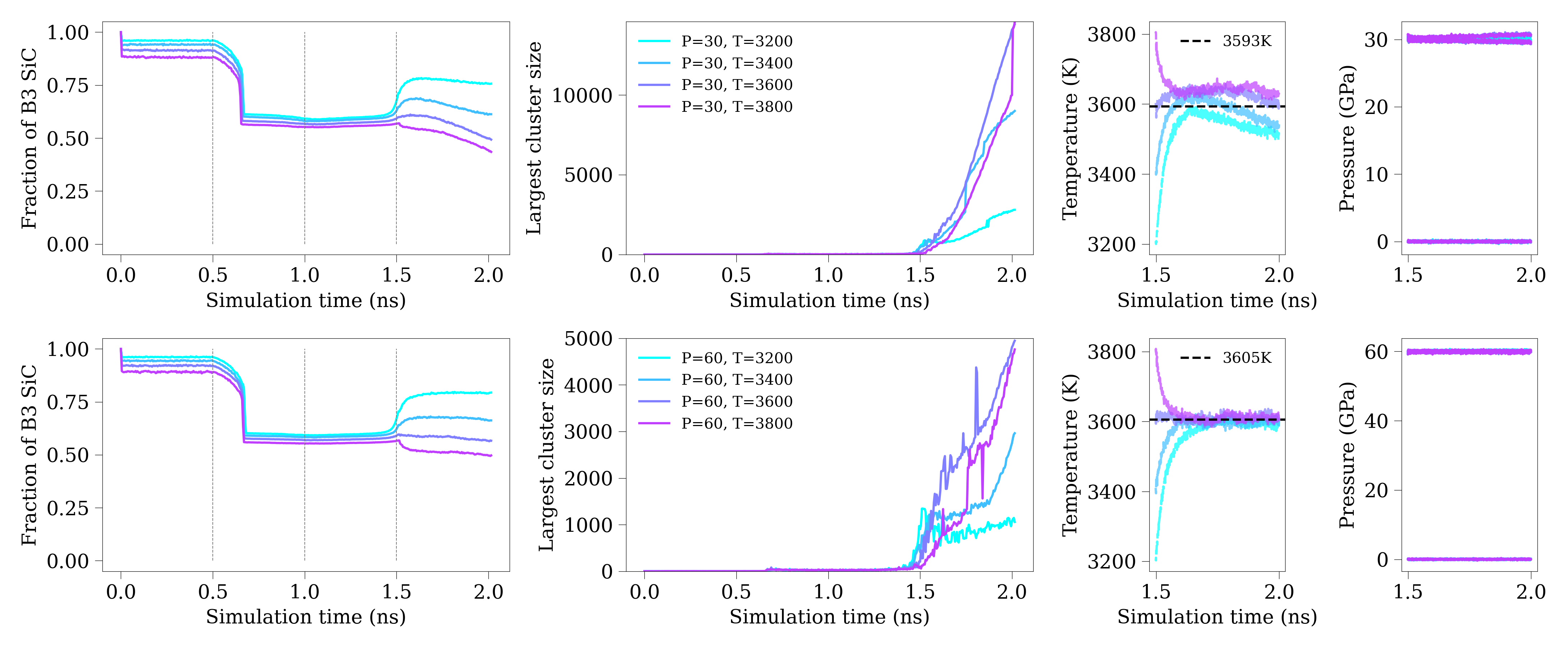}
    \caption{\added{Two-phase MD structural and cluster analysis for simulations of 64,000-atom (top two rows) and 128,000-atom (bottom two rows) at 30 and 60 GPa. Panels from left to right tracks time evolution of: the fraction of B3 SiC, the size of largest carbon cluster, the system temperature, and the full pressure tensor components, including both the diagonal ($P_{xx},P_{yy},P_{zz}$) (around 30 and 60 GPa) and off-diagonal ($P_{xy},P_{xz},P_{yz}$) (around 0 GPa).}}
    \label{fig:large-two-phase}
\end{figure}

\added{To confirm that our 16,000-atom simulations (Fig.~\ref{fig:two_phase_B3}) are sufficiently converged with respect to system size, we compared them to larger two-phase MD simulations of 64,000 and 128,000 atoms at 30 and 60 GPa (Fig.~\ref{fig:large-two-phase}). The calculated phase transition temperatures varied by less than 5 K across these system sizes. This confirms that a 16,000-atom system is large enough to accurately determine the phase transition temperature, while its computational efficiency enables more thorough sampling of additional pressure points along the phase boundary.}

\newpage
\subsection{Uncertainty monitoring}
\begin{figure}[H]
    \centering
    \includegraphics[width=0.9\textwidth]{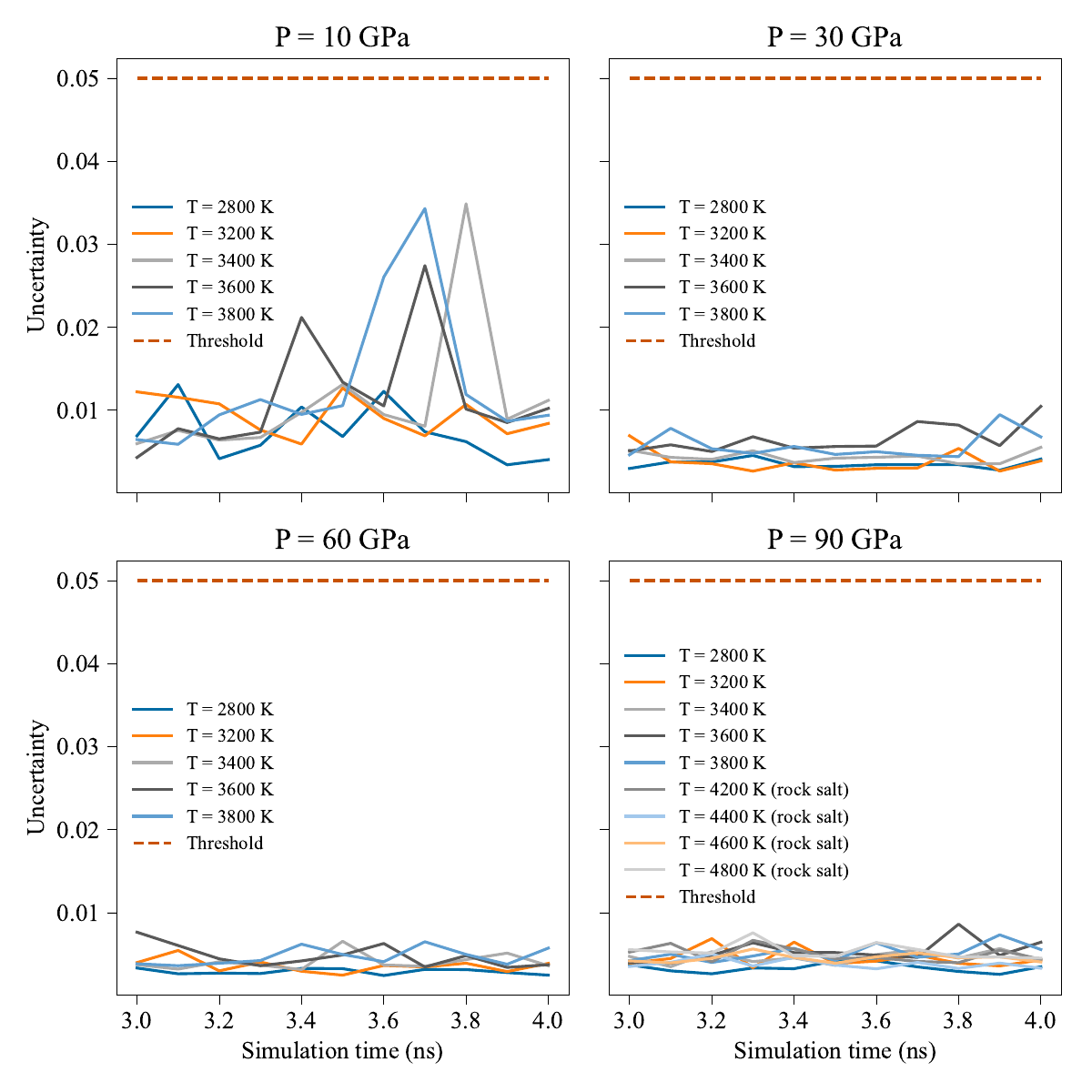}
    \caption{Uncertainty validation of the two-phase MD simulation.}
    \label{fig:my_label}
\end{figure}

\newpage
\section{Other regimes of phase diagram}
\subsection{High-pressure zinc blende - rock salt transition}

\begin{figure}[h]
    \centering
    \includegraphics[width=\textwidth]{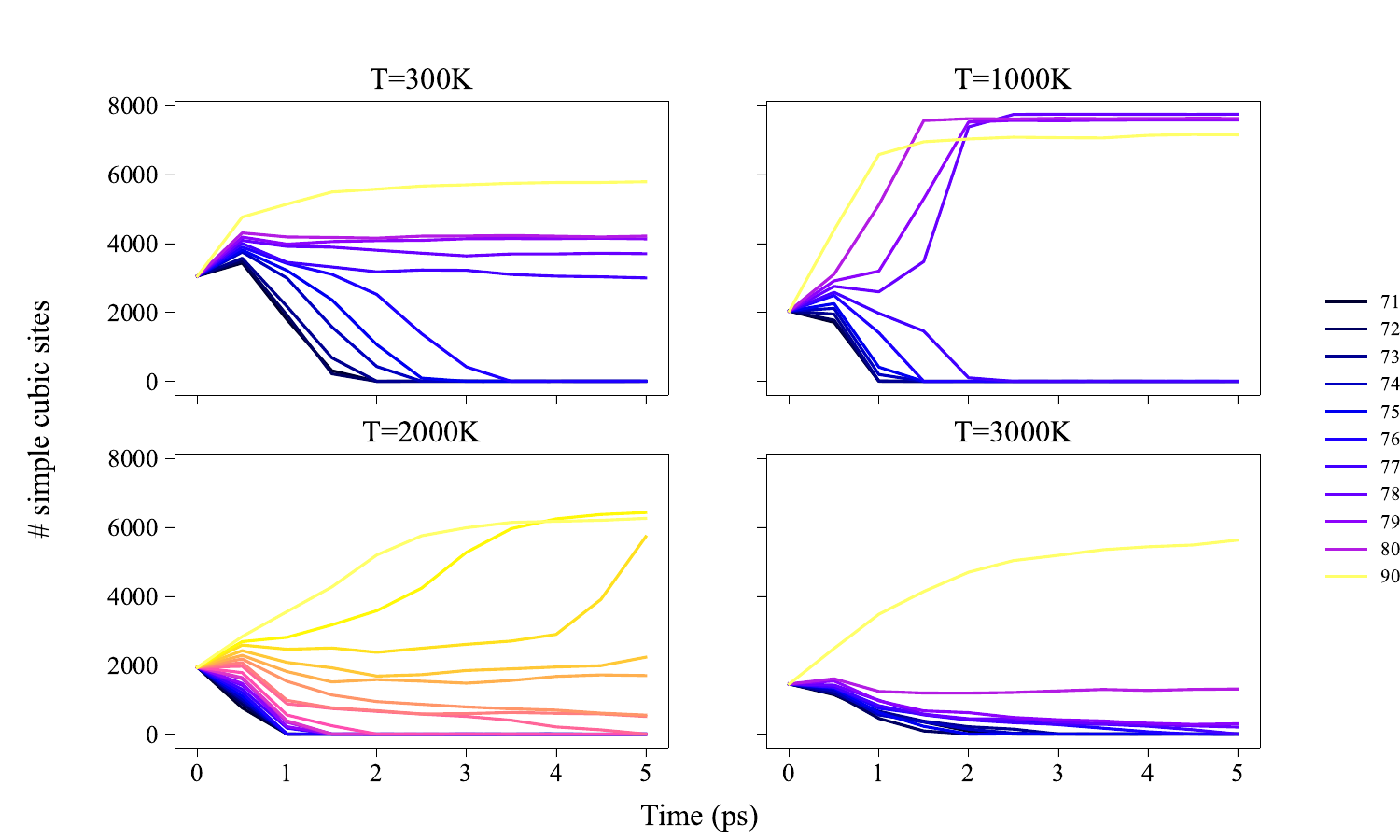}
    \caption{B3$\to$B1 pressure scan plots (pressure unit: GPa).}
    \label{fig:RS_transition}
\end{figure}


\subsection{Gas phase volume change}
\begin{figure}[H]
    \centering
    \includegraphics[width=0.7\textwidth]{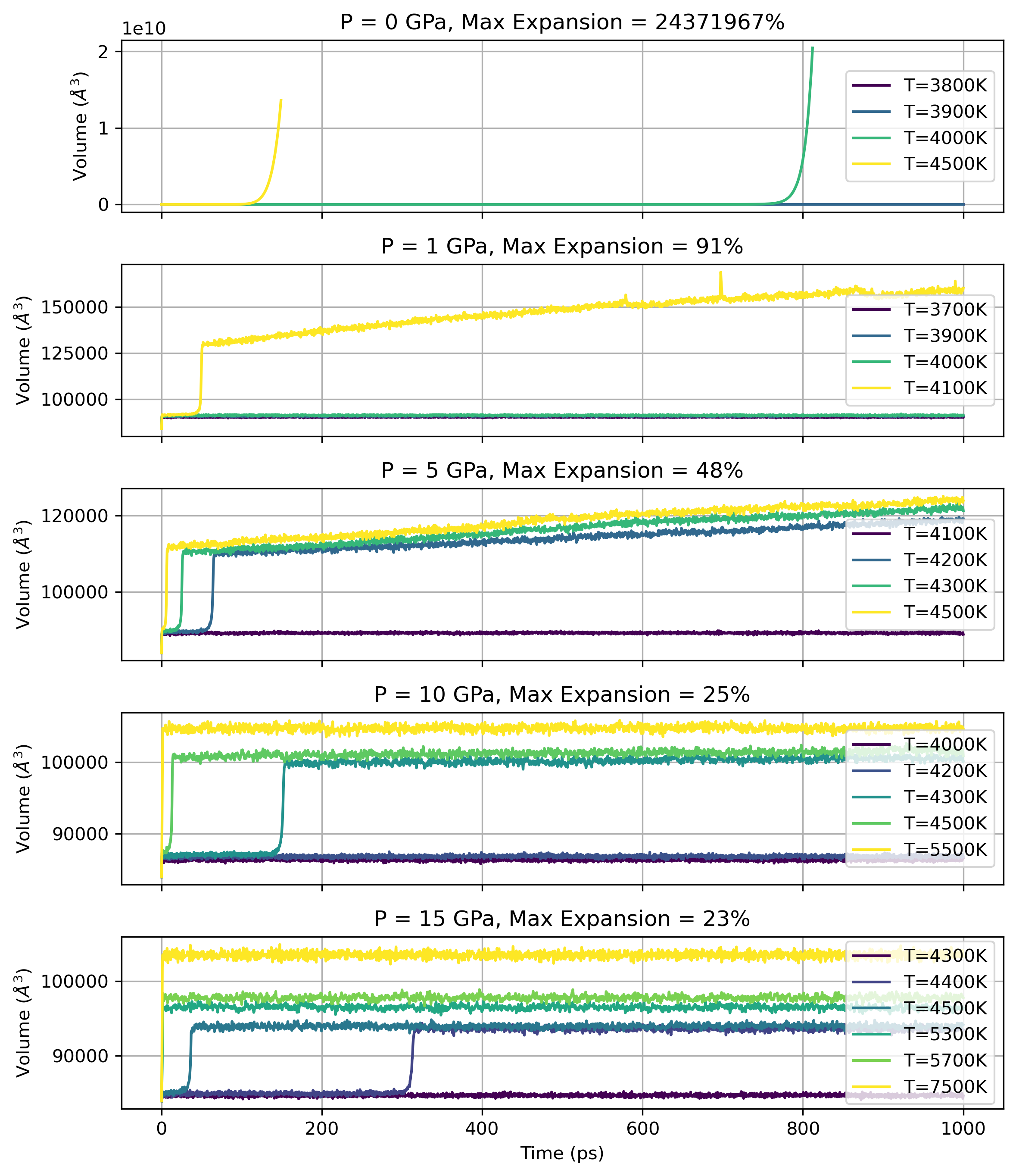}
    \caption{\added{Volume change from MD simulations across pressures and temperatures under $NPT$ ensemble. For each pressure investigated, the percentage volume change is calculated relative to the initial volume at the highest temperature.}}
    \label{fig:volume_change_gas}
\end{figure}

\added{We perform additional $NPT$ MD starting from B3 SiC at various temperatures and at pressures of 0, 1, 5, 10, and 15 GPa. At 0, 1, and 5 GPa, the simulation cell exhibits a sustained, monotonic volumetric expansion once the temperature exceeds approximately 4000, 4100, and 4200 K, respectively, with no subsequent stabilization. This behavior is consistent with progressive sublimation and drift toward a gas/vapor phase rather than a condensed liquid. In contrast, at 10 and 15 GPa, the cell shows an initial transient expansion followed by volume stabilization, indicating the retention of a condensed (non-sublimated) state. These results support a qualitative pressure threshold between 5 and 10 GPa, below which SiC sublimates instead of forming a stable, high-temperature bulk liquid.}

\newpage
\bibliographystyle{naturemag}
\bibliography{main}